%% file: main.tex
\pdfoutput=1

\documentclass[master,english,final]{kaist-ucs}

\usepackage{bookmark}
\usepackage{amsmath,amssymb,amsthm,textcomp,proof,hyperref}
\usepackage{mathpartir}
\usepackage[numbers]{natbib}
\input{coqformat}

\title[korean] {동시성 분리 논리를 사용한 Chase-Lev 덱의 엄밀한 검증}
\title[english]{Formal Verification of Chase-Lev Deque in Concurrent Separation Logic}

\author[korean] {최}{재 민}
\author[korean2] {최}{재민}    
\author[chinese]{崔}{在 珉}
\author[english]{Choi}{Jaemin}

\advisor[major]{강 지 훈}{Jeehoon Kang}{signed}
\advisor[major2]{강지훈}{Jeehoon Kang}{signed}    
\advisorinfo{Professor of Computing} 
\department{CS}{engineering}{a}

\referee[1]{강 지 훈}
\referee[2]{양 홍 석}
\referee[3]{허 기 홍}

\approvaldate{2023}{6}{5}

\refereedate{2023}{6}{5}

\gradyear{2023}

\input{setup}

\begin{document}


   \thesisinfo
   \input{abstract.tex}

    \addtocounter{pagemarker}{1}                 
    \newpage

    \tableofcontents

    \listoftables

    \listoffigures



\chapter{Introduction}
\input{intro.tex}

\chapter{Background: Chase-Lev Deque}
\input{cld.tex}

\chapter{Background: Iris Separation Logic}
\input{iris.tex}

\chapter{Verification of Chase-Lev Deque without Resizing}
\input{verif_simple.tex}

\chapter{Full Verification of Chase-Lev Deque}
\input{verif.tex}

\chapter{Verification in Extended Settings}
\input{extended.tex}

\chapter{Conclusion}
\input{conclusion.tex}

\bibliographystyle{abbrvnat}
\bibliography{main}


\input{ack.tex}


  \label{paperlastpagelabel}     
\end{document}

%% file: coqformat.tex
\usepackage[utf8]{inputenc}
\usepackage[english]{babel}

\usepackage{xcolor}
\usepackage{listings}
\definecolor{dkgreen}{rgb}{0,0.6,0}
\definecolor{ltblue}{rgb}{0,0.4,0.4}
\definecolor{dkviolet}{rgb}{0.3,0,0.5}

\lstdefinelanguage{Coq}{ 
    mathescape=true,
    texcl=false, 
    morekeywords=[1]{Section, Module, End, Require, Import, Export,
        Variable, Variables, Parameter, Parameters, Axiom, Hypothesis,
        Hypotheses, Notation, Local, Tactic, Reserved, Scope, Open, Close,
        Bind, Delimit, Definition, Let, Ltac, Fixpoint, CoFixpoint, Add,
        Morphism, Relation, Implicit, Arguments, Unset, Contextual,
        Strict, Prenex, Implicits, Inductive, CoInductive, Record,
        Structure, Canonical, Coercion, Context, Class, Global, Instance,
        Program, Infix, Theorem, Lemma, Corollary, Proposition, Fact,
        Remark, Example, Proof, Goal, Save, Qed, Defined, Hint, Resolve,
        Rewrite, View, Search, Show, Print, Printing, All, Eval, Check,
        Projections, inside, outside, Def},
    morekeywords=[2]{forall, exists, exists2, fun, fix, cofix, struct,
        match, with, end, as, in, return, let, if, is, then, else, for, of,
        nosimpl, when},
    morekeywords=[3]{Type, Prop, Set, true, false, option},
    morekeywords=[4]{pose, set, move, case, elim, apply, clear, hnf,
        intro, intros, generalize, rename, pattern, after, destruct,
        induction, using, refine, inversion, injection, rewrite, congr,
        unlock, compute, ring, field, fourier, replace, fold, unfold,
        change, cutrewrite, simpl, have, suff, wlog, suffices, without,
        loss, nat_norm, assert, cut, trivial, revert, bool_congr, nat_congr,
        symmetry, transitivity, auto, split, left, right, autorewrite},
    morekeywords=[5]{by, done, exact, reflexivity, tauto, romega, omega,
        assumption, solve, contradiction, discriminate},
    morekeywords=[6]{do, last, first, try, idtac, repeat},
    morecomment=[s]{(*}{*)},
    showstringspaces=false,
    morestring=[b]",
    morestring=[d],
    tabsize=3,
    extendedchars=false,
    sensitive=true,
    breaklines=false,
    basicstyle=\small,
    captionpos=b,
    columns=[l]flexible,
    identifierstyle={\ttfamily\color{black}},
    keywordstyle=[1]{\ttfamily\color{dkviolet}},
    keywordstyle=[2]{\ttfamily\color{dkgreen}},
    keywordstyle=[3]{\ttfamily\color{ltblue}},
    keywordstyle=[4]{\ttfamily\color{dkblue}},
    keywordstyle=[5]{\ttfamily\color{dkred}},
    stringstyle=\ttfamily,
    commentstyle={\ttfamily\color{dkgreen}},
    literate=
	  {\#}{{\code{\#}}}1
      {<- }{{\code{<- }}}1
	  {:}{{\code{:}}}1
	  {:= }{{\code{:= }}}1
	  {!}{{\code{!}}}1
	  {;}{{\code{;}}}1
}[keywords,comments,strings]

%% file: setup.tex
\usepackage{iris,pftools}

\theoremstyle{definition}

\usepackage{scalerel}
\DeclareMathOperator*{\bigsep}{\scalerel*{\ast}{\sum}}
\newcommand{\AU}[3]{\textlog{AU}_{{#1}, {#2}}(#3)}
\newcommand{\au}{\textlog{AU}}

\def\msf#1{\ensuremath{\mathsf{#1}}}
\newcommand{\code}[1]{{\texttt{#1}}}

\usepackage{stmaryrd}
\newcommand{\ghostVar}[3]{\ownGhost{}{#1 \Mapsto^{#2} #3}}
\newcommand{\natauth}[2]{\ownGhost{#1}{\authfull \textlog{MN}\ #2}}
\newcommand{\natfrag}[2]{\ownGhost{#1}{\authfrag \textlog{MN}\ #2}}
\newcommand{\mapauth}[2]{\ownGhost{#1}{\authfull \textlog{Map}\ #2}}
\newcommand{\mapfrag}[3]{\ownGhost{#1}{\authfrag \textlog{Map}\ #2 \mapsto ^{\always} #3}}
\newcommand{\mapfraghalf}[3]{\ownGhost{#1}{\authfrag \textlog{Map}\ #2 \mapsto ^{1/2} #3}}

\usepackage{skak}
\setboardfontsize{12pt}
\newcommand{\dqstauth}{\BlackKingOnWhite}
\newcommand{\dqstfrag}{\WhitePawnOnWhite}
\newcommand{\dqstarch}{\WhiteKingOnWhite}
\newcommand{\dqstauthg}[2]{\ownGhost{#1}{\dqstauth({#2})}}
\newcommand{\dqstfragg}[2]{\ownGhost{#1}{\dqstfrag({#2})}}
\newcommand{\dqstauthf}[2]{\dqstauth({#2})^{#1}}
\newcommand{\dqstfragf}[2]{\dqstfrag({#2})^{#1}}
\newcommand{\dqstarchf}[2]{\dqstarch({#2})^{#1}}

\usepackage{mathrsfs}
\newcommand{\ownDeque}{\textlog{OwnDeque}}
\newcommand{\deque}{\textlog{Deque}}
\newcommand{\dequeInv}{\textlog{DequeInv}}
\newcommand{\preservation}[2]{\textlog{Preservation}_{{#1},{#2}}}
\newcommand{\dequeLocal}{\textlog{DequeLocal}}
\newcommand{\circleInv}{\textlog{CircleInv}}
\newcommand{\TS}{{\mathscr{T}}}

\newcommand{\ownDequeg}{\ownDeque^\gamma}
\newcommand{\dequeg}{\deque^\gamma}
\newcommand{\dequeInvg}{\dequeInv^\gamma}

\definecolor{StringRed}{rgb}{.58,0.06,0.06}
\definecolor{CommentGreen}{rgb}{0.0,0.55,0.3}
\definecolor{KeywordBlue}{rgb}{0.0,0.3,0.55}
\definecolor{LinkColor}{rgb}{0.55,0.0,0.3}
\definecolor{CiteColor}{rgb}{0.55,0.0,0.3}
\definecolor{HighlightColor}{rgb}{0.0,0.0,0.0}
\definecolor{Protected}{rgb}{.20,.50,.70} 
\definecolor{Managed}{rgb}{.64,.40,.20} 
\definecolor{ProofShade}{cmyk}{0,0,0,0.05}

\newcommand*\Managed{\msf{\color{Managed}Managed}}
\newcommand*\Shield{\msf{HPSlot}}
\newcommand*\Protected{\msf{\color{Protected}Protected}}

\newcommand{\SN}{\textlog{sn}}
\newcommand{\SW}{\textlog{sw}}
\newcommand{\snseen}{\sqsupseteq_\SN}
\newcommand{\swseen}{\sqsupseteq_\SW}
\newcommand{\roseen}{\sqsupseteq_\RO}
\newcommand{\CAS}{\textlog{cas}}
\newcommand{\RO}{\textlog{ro}}

\usepackage{pifont}
\newcommand{\yes}{\textcolor{CommentGreen}{\ding{51}}}%
\newcommand{\no}{\textcolor{StringRed}{\ding{55}}}%
\newcommand{\yesno}{\textcolor{olive}{\ding{115}}}%

%% file: abstract.tex
\begin{summary}      
Chase-Lev 덱은 멀티프로세서 스케줄링에서 효율적으로 부하를 분산시키는 데 사용되는 동시성 자료 구조이다.
이는 다음과 같은 작업 훔치기 기법을 지원한다.
각 스레드는 작업 저장소로서 Chase-Lev 덱을 하나씩 소유하고, 할 작업이 없어진 스레드는 다른 스레드의 작업 저장소로부터 작업을 훔쳐 대신 실행한다.
하지만 멀티프로세서 환경을 비롯한 모든 소프트웨어에서 버그의 위험이 내재되어 있기 때문에, 프로그램이나 자료 구조가 올바르게 동작함을 엄밀하게 증명하는 것이 중요하다.
현재까지 알려진 바로는 Chase-Lev 덱의 엄밀한 검증 연구 중
(1) 증명에서 믿고 넘어가야 하는 요소가 가능한 한 작고,
(2) 현실적이면서 제약이 없는 구현을 사용하며,
(3) 강한 명세를 증명한
사례는 없었다.

본 논문에서는 이러한 한계를 해결하기 위해 동시성 분리 논리를 사용하여 Chase-Lev 덱의 엄밀한 검증을 제시한다.
이 검증은 Coq 증명 보조 도구를 사용하여 작성되었으며,
검증된 구현은 현실적이면서 작업의 수에 제약이 없다.
또한 동시성 자료 구조의 강한 명세로 흔히 인정되는 linearizability를 명세로 사용한다.
따라서 본 연구의 검증은 위의 세 조건을 모두 충족한다.
추가로, 검증 작업을 확장하여 메모리 재활용 기법을 사용하는 구현을 검증하고, 느슨한 메모리 모델에서 Chase-Lev 덱을 검증하기 위한 토대를 마련하였다.
\end{summary}
   
\begin{Korkeyword}
동시성, 엄밀한 검증, 프로그램 논리, 자료 구조, 분리 논리, Chase-Lev 덱
\end{Korkeyword}

\begin{abstract}
Chase-Lev deque is a concurrent data structure designed for efficient load balancing in multiprocessor scheduling.
It employs a work-stealing strategy, where each thread possesses its own work-stealing deque to store tasks, and idle threads steal tasks from other threads.
However, given the inherent risk of bugs in software, particularly in a multiprocessor environment, it is crucial to formally establish the correctness of programs and data structures.
To our knowledge, no formal verification work for the Chase-Lev deque has met three key criteria:
(1) utilizing a minimal trusted computing base,
(2) using a realistic and unrestricted implementation, and
(3) proving a strong specification.

In this thesis, we address this gap by presenting the formal verification of the Chase-Lev deque using a concurrent separation logic.
Our work is mechanized in the Coq proof assistant,
and our verified implementation is both realistic and unbounded in terms of the number of tasks it can handle.
Also, we adopt linearizability as the specification, as it is widely recognized as a strong specification for concurrent data structures.
Consequently, our work satisfies all three aforementioned criteria for formal verification. 
Additionally, we extend our verification to support safe memory reclamation, and provide a basis for verifying the Chase-Lev deque in the relaxed memory model.
\end{abstract} 
     
\begin{Engkeyword}
Concurrency, formal verification, program logic, data structure, separation logic, Chase-Lev deque
\end{Engkeyword}

%% file: intro.tex
\label{chap:intro}
\section{Background: Chase-Lev Deque for Efficient Work-stealing}

\begin{figure}[t]
  \includegraphics[width=9cm]{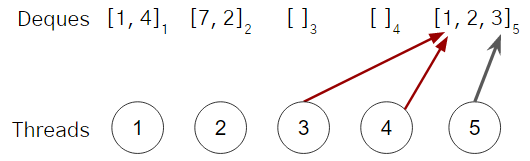}
  \centering
  \caption{A demonstration of work-stealing for 5 threads.}
  \label{fig:stealing}
\end{figure}


High throughput is a significant obstacle when it comes to managing multiple processors concurrently.
To optimize this objective, efficient load balancing is required.
The aim is to allocate tasks evenly, ensuring that certain threads are not overwhelmed while others remain idle.
However, due to various constraints such as dependencies and unknown execution times, it is challenging to distribute tasks effectively from the get-go.

Work-stealing \citep{bl} tackles this problem by dynamically changing the work distribution.
In this approach, when a thread becomes idle, it actively searches for tasks in the task pool of other threads and ``steals'' a task to execute, ensuring that idle threads remain productive.
Each thread maintains its own task storage called a \textit{work-stealing deque}, containing the tasks assigned to it.
When a thread has a work to perform, it removes a task from its deque\footnote{In this paper, we may refer to work-stealing deques as just deques, although they don't provide all the APIs of a regular deque.}.
Also, when a thread becomes idle but its deque is empty, it steals a task from another thread's deque.
This is illustrated in \autoref{fig:stealing}, where thread 5 is trying to remove task 3 from its own deque, while thread 3 and 4 are trying to steal task 1 from thread 5's deque.

Numerous work-stealing deques have been developed for efficient work-stealing
\citep{abp, cl, cl-weakmem, fence-free, low-synch, acar, lace, cilk, bwos}.
Among these, the work presented by Chase and Lev \citep{cl}, known as the \textit{Chase-Lev deque}, stands out as a popular high-performance and realistic design.
It achieves fine-grained concurrency by avoiding the use of locks and allows for an unlimited number of elements in a single deque.
The Chase-Lev deque is widely used in various real-world concurrency frameworks \citep{crossbeam, taskflow}, further demonstrating its practicality and effectiveness.

\section{Problem: Lack of Formal Verification for Chase-Lev Deque}

\subsection{Necessity of Formal Verification}

However, the multiprocessor environment presents additional complexities and challenges compared to single-threaded settings, making it more prone to errors.
In the context of work-stealing, multiple threads may concurrently attempt to steal the same task from the same deque, leading to collisions.
At the same time, the owner thread may also try to insert or remove its tasks, potentially colliding with the stealers.
A proper synchronization is required to handle this situation:
an incorrect implementation can result in various bugs, such as tasks being executed multiple times, removed in an undesirable order, or not inserted into the deque.


The complexity of the situation further intensifies in modern architectures and compilers.
Most of this thesis assumes the \textit{sequentially consistent (SC) memory model}, where instructions within each thread are executed in order.
However, modern architectures follow the \textit{relaxed memory model}, which allows for out-of-order execution of instructions for optimization purposes.
While programmers can impose certain ordering constraints to ensure proper synchronization, it must be done with caution.
Overly strict ordering can degrade the performance, while overly weak ordering can introduce bugs.
It is worth noting that even the implementation of Chase-Lev deque in relaxed-memory, written and peer-reviewed by experts \citep{cl-weakmem}, was found to contain a bug \citep{cdschecker}.

Given the complexity of real-world architecture and programs, how can we ensure the absence of bugs?
While program testing can help identify unexpected bugs, it does not guarantee the lack of them.
As such, it is not uncommon to see bug reports in heavily tested commercial programs.
This gives rise to the significance of \textit{formal verification}, the process of specifying and proving the correctness of programs using formal methods.
By providing a rigorous and systematic approach to verifying program correctness, formal verification helps mitigate the risk of bugs and ensures a higher level of reliability in software.

\subsection{Prior Works on Chase-Lev Deque Verification}

To the best of our knowledge, there has not been a foundational, realistic, and strong verification of Chase-Lev deque.
Specifically, the following are the properties of interest for formal verification: \begin{itemize}
	\item \textbf{Foundational verification}:
	the verification process should be mechanized in proof assistants like Coq \citep{coq}, allowing for a minimal trusted computing base.
	While model checkers offer the advantage of automation, their credibility relies on trusting the theory and implementation of the whole model checking program. 
	On the other hand, proof assistants only require trust in the proof checker and not the underlying logic or framework.
	\item \textbf{Realistic implementation}:
	the implementation under verification should not be simplified to the extent of being restrictive or unrealistic.
	It is crucial to capture the essential complexities and behaviors of the actual Chase-Lev deque to ensure that the verification results hold in practical scenarios.
	\item \textbf{Strong specification}:
	the verified specifciation should be strong enough to allow clients to utilize the data structure in various ways.
	In the SC memory model, \textit{linearizability} \citep{linearizability} is a de facto standard for the correctness of concurrent data structures.
	For the relaxed memory model, the Compass framework \citep{compass} has recently been developed to provide a strong specification.
\end{itemize}

\citet{cl-weakmem} implemented Chase-Lev deque in the relaxed memory model and provided a proof of its correctness.
However, the proof was conducted using pen-and-paper, which is not sufficient for a foundational verification.
Pen-and-paper proofs are prone to human errors and require thorough review compared to machine-checked proofs.
Additionally, although the verified specification is strong with regard to the SC memory model, it does not fully capture the relaxed memory behavior such as the synchronization between clients of the deque.
Moreover, the lack of modularity in the specification poses challenges when verifying programs that utilize the Chase-Lev deque, as it is not straightforward to incorporate and reason about its specification in a modular manner.

\citet{cl-refine} verified the linearizability of the Chase-Lev deque in the SC memory model using the Civl verifier \citep{civl}.
As such, it is the first machine-checked proof of linearizability for the Chase-Lev deque.
However, it is not foundational: Civl is a complex proof system, and its correctness should be trusted for the basis of the verification.
Furthermore, the implementation in the verification assumes an infinitely large array, which is unrealistic in practice.
It is crucial to note that this is not a minor simplifying assumption, as the synchronization and reasoning required in the full implementation are significantly more complex than with this simplification.

\citet{genmc} developed a model checker for C programs in configurable memory models, and used it to verify the Chase-Lev deque in a relaxed memory setting as a benchmark.
While it has an advantage of verifying a C implementation in the relaxed memory model, it does not meet any of the three criteria of interest.
First, it relies on the model checker as a trusted computing base.
Second, although it uses a finite array, the capacity of the deque is limited \citep{genmc-bench-cl}, introducing the same simplification issue as assuming an infinite array.
Finally, linearizability or other strong specifications were not verified, but only some weaker guarantees such as safety.


\begin{table}
  \centering
  \begin{tabular}{|c|c|c|c|c|c|c|}
    & Machine- & Foundational & Realistic & Strong & Relaxed & Real-world
    \\ & checked & & impl. & spec & memory & language
    \\ \hline
    \citet{cl-weakmem} & \no & \no & \yes & \yesno & \yes & \yes
    \\ \hline
    \citet{cl-refine} & \yes & \no & \no & \yes & \no & \no
    \\ \hline
    \citet{genmc} & \yes & \no & \no & \no & \yes & \yes
    \\ \hline
    Our work & \yes & \yes & \yes & \yes & \yesno & \no
  \end{tabular}
  \caption{Comparison of Chase-Lev deque verification.}
  \label{table:comparison}
\end{table}

The comparison of Chase-Lev deque verification approaches is summarized in \autoref{table:comparison}.
In addition to the the criteria of our interest plus machine-checked verification (as a subcriterion of foundational verification),
we also compare them with regard to verification in the relaxed memory model, and the use of real-world languages like C.

\subsection{Verification Challenges}

Recent advancements in formal verification have enabled the foundational verification of strong specifications for various concurrent data structures,
including those utilized in industrial projects \citep{folly, dartino}, as well as those in relaxed memory \citep{gps, compass}. 
However, the verification of the Chase-Lev deque poses unique challenges.
Its complexity lies in the intricacies of synchronization, to the extent that it is nontrivial to comprehend its correctness even intuitively.

The Chase-Lev deque uses a dynamic circular array which automatically resizes on overflow.
The contents of the deque are represented by a circular slice of this array.
The deque also maintains two integer indices, namely \textit{top} and \textit{bottom}.
These indices denote the starting and ending positions of the slice, respectively.
The owner of the deque may insert or remove elements (tasks) from the bottom end of the deque, and the stealers may remove elements from the top end.

An obvious source of conflict arises when multiple threads attempt to steal the same element from the top end of the deque.
In addition, if only one element remains, the owner thread may also attempt to pop that element, potentially joining the conflict.
This situation is handled by a CAS operation on the top index, and the verification can proceed by just doing a case analysis on whether the CAS succeeded.

However, the synchronization involved in the Chase-Lev deque extends beyond simple CAS operations.
Verifying its correctness requires intricate reasoning about the array modifications.
One source of complication is that the value to steal is determined before CAS-ing.
A steal operation involves CAS-ing the top index, but before that, the stealer must read the array to remember the value it intends to steal
(the reason will be elaborated in \autoref{chap:cld}).
After a successful CAS, the value remembered will be returned.
Then why does it work correctly, despite the owner being able to pop everything and push a new element in the meantime, or wrap around and overwrite the slot by pushing?
To answer this question, we should be able to establish that certain values are preserved during a specific period of execution.

Another challenge arises from the fact that the owner can replace the array while a stealer is in the process of stealing.
Consider a scenario where a stealer reads the address of the array but then gets stalled, and the owner replaces the array while pushing new elements.
It is even possible to replace the array multiple times by pushing a significant number of elements.
Eventually, the stealer resumes execution but reads a value from the old, replaced array.
Surprisingly, despite the array being replaced, the stealer can still successfully complete the operation and affect the deque's state.
This is in contrast to other data structures like Harris' linked list \citep{harris-list}, where an operation fails and restarts if it detects a detached node.
To account for this situation in the Chase-Lev deque, the verification process must reason not only about the current array but also about all past arrays, and establish some form of linkage between them.

\section{Our Solution: Foundational Verification of Chase-Lev Deque}

This thesis presents the first full foundational verification of Chase-Lev deque using the Iris separation logic \citep{iris, iris-1}.
The verification is mechanized in the Coq proof assistant \citep{coq}, using the mechanization for Iris.
Our verified implementation of the deque is both realistic and unbounded, as it utilizes a finite circular array that dynamically resizes upon overflow.
Moreover, we establish the linearizability of the Chase-Lev deque, which provides a strong specification in the SC memory model.
Thus, our verification satisfies all the criteria of foundational verification, realistic implementation, and strong specification.
Specifically, we make the following contributions:
\begin{itemize}
  \item In \autoref{chap:verif-simple}, we present the verification of the Chase-Lev deque without considering array resizing.
  \item In \autoref{chap:verif}, we extend the verification to encompass the full implementation of the Chase-Lev deque, which includes array resizing.
  \item In \autoref{chap:extended}, we explore the extension of our verification approach to different settings. Specifically, we verify Chase-Lev deque under safe memory reclamation, and provide a basis for verification in the relaxed memory model.
\end{itemize}

The other chapters are organized as follows.
In \autoref{chap:cld}, we describe the implementation of the Chase-Lev deque, and explain the intuition behind its correctness.
In \autoref{chap:iris}, we give a brief introduction to the Iris separation logic, focusing on the features relevant to our work.
In \autoref{chap:conclusion}, we summarize our results, and present related and future works.
All our results are mechanized in Coq, and the mechanization for \autoref{chap:verif-simple} and \autoref{chap:verif} are available in the following link: \url{https://github.com/kaist-cp/chase-lev-verification}.
The mechanization for \autoref{chap:extended} will be published in the future along with the corresponding papers.

%% file: cld.tex
\label{chap:cld}

\section{The APIs of Work-stealing}

Recall from \autoref{fig:stealing}: in a scheduling scheme that supports work-stealing, each thread maintains its own work-stealing deque containing its assigned tasks. The owner of a deque can push a newly assigned task to its deque, or pop a task from its deque and start executing it. Other threads, which we will call \textit{stealers}, can steal a task from the same deque so that they can execute it instead. This helps balancing out the workload, and preventing some threads to become idle while other threads are overloaded with tasks. Technically, the owner can also steal a task from its own deque, although this breaks the owner's LIFO behavior.

A work-stealing deque provides the following deque-like interface:
\begin{itemize}
  \item \textbf{Push} inserts a task at the bottom end of the deque. This method can only be called by the owner of the deque.
  \item \textbf{Pop} tries to remove a task from the bottom end of the deque. This method can only be called by the owner of the deque. The attempt may fail if the deque is empty or it clashes with other threads' steals.
  \item \textbf{Steal} tries to remove a task from the top end of the deque. This method can be called by any thread. The attempt may fail if the deque is empty or it clashes with other threads' steals or the owner's pop.
\end{itemize}

\section{Structure of Chase-Lev Deque}

\begin{figure}[t]
  \includegraphics[width=8cm]{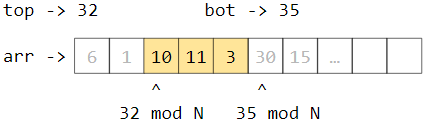}
  \centering
  \caption{A Chase-Lev deque with the content $[10, 11, 3]$.}
  \label{fig:structure}
\end{figure}

Now we go over the details of Chase-Lev deque. The structure of the deque is illustrated in \autoref{fig:structure}. It is implemented as a circular array, along with top and bottom indices. Being a circular array, the indexing is done modulo the array's size. If the top index, bottom index, and array are $t$, $b$, and $arr$, respectively, then the contents of the deque is represented by the half-open circular slice $[t, b)$, i.e. $[arr[t \text{ mod } N], arr[(t+1) \text{ mod } N], \cdots, arr[(b-1) \text{ mod } N]]$ where $N = |arr|$.
Pushing a value to the deque amounts to writing the value at the bottom index of the array and incrementing the bottom. Similarly, a successful pop decrements the bottom index and returns the value at the new bottom index (except in a corner case we will soon discuss); a successful steal does the same but uses the top index.

Additionally, the owner resizes the array when it tries to push but the array is full. To do this, the owner allocates a larger array, copies the values from top to bottom in a way such that modulo indexing gives the same value, and substitutes the array with the new one.

Since all array accesses are done modulo its size, we will denote $arr[i]$ as a shorthand for $arr[i \text{ mod } |arr|]$.

\section{Implementation}

The implementation of Chase-Lev deque in HeapLang, a language provided by Iris, is presented in this section.

\begin{figure}
  \noindent\begin{minipage}{\textwidth}
  \begin{lstlisting}[frame=tlrb,language=Coq]{Name}
Definition new_wsdeque :   val := λ:   "sz",
  let: "array" := AllocN "sz" #0 in
  (ref ("array", "sz"), ref #1, ref #1). (* array+size, top, bot *)
    
Definition arr :  val := λ:  "deque", Fst (Fst "deque").
Definition top :  val := λ:  "deque", Snd (Fst "deque").
Definition bot :  val := λ:  "deque", Snd "deque".
Definition access :  val := λ:  "arr" "i" "n", "arr" +ₗ ("i" `rem` "n").
  \end{lstlisting}
  \end{minipage}\hfill
  \caption{The initialization and field access functions.}
  \label{fig:new}
\end{figure}

The initialization and field access functions are given in \autoref{fig:new}. A deque is represented as a tuple of circular array, top index, and bottom index. The top and bottom indices start from 1 instead of 0 just for a technical reason. Also, the size is stored along with the array itself so that threads can index into it modulo its size. \verb|arr|, \verb|top|, and \verb|bot| are used as shorthands for field accesses. \verb|access| is used to access a slot of the array: it takes a circular array $arr$, index $i$, and the size of the array $n$, and returns $arr[i]$. 

\begin{figure}
  \noindent\begin{minipage}{.45\textwidth}
  \begin{lstlisting}[frame=tlrb,language=Coq]{Name}
Definition grow :  val := λ:  "circle" "t" "b",
  let: "sz" := Snd "circle" in
  let: "nsz" := #2 * "sz" in
  let: "narr" := AllocN "nsz" #0 in
  (* copy from Fst "circle" to "narr" for
    the indices in [t, b), omitted here *)
  ("narr", "nsz").
  \end{lstlisting}
  \end{minipage}\hfill
  \begin{minipage}{.45\textwidth}
  \begin{lstlisting}[frame=tlrb,language=Coq]{Name}
Definition push :  val := λ:  "deque" "v",
  let: "b" := !(bot "deque") in
  let: "t" := !(top "deque") in
  let: "circle" := !(arr "deque") in
  let: "sz" := Snd "circle" in
  (if: "t" + "sz" ≤ "b" + #1
    then arr "deque" <- grow "circle" "t" "b"
    else #()
  ) ;;
  let: "circle'" := !(arr "deque") in
  let: "sz'" := Snd "circle'" in
  (access (Fst "circle'") "b" "sz'") <- "v" ;;
  bot "deque" <- "b" + #1.  
  \end{lstlisting}
  \end{minipage}
  \caption{The push function.}
  \label{fig:push}
\end{figure}

Now we describe the implementation of the three main APIs. The push function is given in \autoref{fig:push}.
\verb|push| uses the function \verb|grow| to replace the deque's array. It takes a circular array $circle$, top $t$, and bottom $b$, and allocates and returns a new circular array $circle'$ with double the size of the original one, such that they have the same values from $t$ to $b$ modulo their sizes: $\forall i, t \leq i < b \implies circle[i] = circle'[i]$. We skip the implementation detail here; refer to the link provided in \autoref{chap:intro}.

To push a value $v$, we start by reading bottom ($b$), top ($t$), and array ($arr$) along with its size ($sz$). If the array is full, i.e. $sz \leq b - t + 1$, the array is replaced by the \verb|grow| function which returns a new circular array. Then, we read the array ($arr'$) and its size ($sz'$) again since they might have been changed, write the value $v$ at $arr'[b]$, and increment the bottom.

\begin{figure}
  \noindent\begin{minipage}{.5\textwidth}
  \begin{lstlisting}[frame=tlrb,language=Coq]{Name}
Definition pop : val := λ: "deque",
  let: "b" := !(bot "deque") - #1 in
  let: "circle" := !(arr "deque") in
  let: "sz" := Snd "circle" in
  bot "deque" <- "b" ;;
  let: "t" := !(top "deque") in
  if: "b" < "t" then
    (* empty pop *)
    bot "deque" <- "t" ;; NONE
  else let: "v" :=
    !(access (Fst "circle") "b" "sz") in
  if: "t" < "b" then
    (* normal case *)
    SOME "v"
  else let: "ok" :=
    CAS (top "deque") "t" ("t" + #1) in
  bot "deque" <- "t" + #1 ;;
  if: "ok" then SOME "v" (* popped *)
  else NONE. (* stolen *)  
  \end{lstlisting}
  \end{minipage}\hfill
  \begin{minipage}{.45\textwidth}
  \begin{lstlisting}[frame=tlrb,language=Coq]{Name}
Definition steal : val := λ: "deque",
  let: "t" := !(top "deque") in
  let: "b" := !(bot "deque") in
  let: "circle" := !(arr "deque") in
  let: "sz" := Snd "circle" in
  if: "b" ≤ "t" then
    (* no chance *)
    NONE 
  else let: "v" :=
    !(access (Fst "circle") "t" "sz") in
  if: CAS (top "deque") "t" ("t" + #1)
  then SOME "v" (* success *)
  else NONE. (* fail *)    
  \end{lstlisting}
  \end{minipage}
  \caption{The pop and steal functions.}
  \label{fig:popsteal}
\end{figure}

Next, the pop and steal function is given in \autoref{fig:popsteal}.
We describe \verb|steal| function first, because we have to understand it first to discuss \verb|pop|. If $b \leq t$, there is nothing to steal. Otherwise, we read $arr[t]$ and then attempt to steal by CAS-ing the top to $t+1$. Upon a successful CAS, the value we read earlier is returned. It is important to read $arr[t]$ \textit{before} CAS-ing: after CAS succeeds, the owner may wrap around and push a new element at $arr[t + sz]$. Then it is too late to read $arr[t]$ because it was overwritten. On the other hand, it is correct to read it before CAS-ing, because $arr[t]$ is guaranteed to stay the same until the successful CAS. We will discuss this point in more detail in the next section.

The last function to discuss is \verb|pop|, which may look peculiar. The explanation is as follows:
\begin{itemize}
  \item We start by immediately decrementing the bottom, as if the ``bottom element'' is already popped, even though the deque may be empty. If the deque \textit{is} empty, we increment it back.
  
  It is actually mandatory to decrement the bottom before reading the top. To see why, suppose the owner does not decrement the bottom ($b$) and read the top ($t$), then learn $b > t$ so we enter the ``normal case'' branch. But right after reading the top, stealers come in and steal every element. Now the owner resumes and pops the bottom element, not realizing that it was already stolen, resulting in an incorrect behavior since then multiple threads would execute the same task at the same time. Any attempt to read the bottom and top again in the normal case would suffer from the same problem of stealing after reading. We instead decrement the bottom prematurely, protecting the bottom element from the stealers.
  
  Of course, this does not prevent the stealers from stealing the bottom element before even decrementing the bottom. Fortunately, this case is safe because then $t$ would be large enough that we enter the ``empty pop'' branch.
  \item If there are more than one elements in the deque, we enter the ``normal case'' branch where the bottom element is simply returned. As we saw earlier, this element is safe from concurrent steals.
  \item If there is only one element, it is incorrect to just return the bottom element. Instead, we CAS the top just like stealing. The reason is because there is a potential conflict with concurrent stealers. Specifically, suppose a stealer reads the top and bottom before the owner decrements the bottom. Then the stealers enter the normal case where the CAS on top is attempted. At the same time, the owner starts to pop and notices that there is only one element. At this point, if the stealer succeeds the CAS, the owner should not be able to pop the only element since it is being stolen. This conflict is resolved by joining the stealers and CAS-ing the top. Despite popping from the top, this does not break the owner's LIFO behavior since there is no difference from the client's viewpoint.
\end{itemize}

\section{Intuition}

Now we discuss a few observations that will be used in verifying Chase-Lev deque. First, only the owner can modify the bottom and the array. This is because \verb|steal| does not do so. As a result, the owner can completely keep track of these two fields. This property is especially crucial in \verb|grow| since the contents are copied over multiple steps, and in \verb|pop| since it involves a complex reasoning on the bottom index.

Next, the top can only increase. This is because the only way to change the top is by reading its value $t$ and CAS-ing it to $t+1$. As a corollary, only one of the CAS attempts from $t$ to $t+1$ can ever succeed.

Finally, if the top $t$ and bottom $b$ satisfies $t < b$, the inequality stays true and $arr[t]$ is preserved until $t$ increases. This is because the only way to remove the element at the top (if it exists), either from \verb|pop| or \verb|steal|, is by CAS-ing the top. This property holds even if $arr$ has already been replaced, because the values in the old array never get overwritten. This is why it is safe for \verb|steal| to read $arr[t]$ before CAS-ing the top even if the owner replaces the array in the meantime.

%% file: iris.tex
\label{chap:iris}
Using the intuitive properties discussed in \autoref{chap:cld}, we can informally explain why Chase-Lev deque ``works correctly'', e.g. no elements are removed twice, the stealers remove the elements in FIFO order, and so on. However, we seek foundational verification; we need a way to formally express the specification and the reasoning.

In this chapter, we introduce Iris, a framework for concurrent separation logic \cite{iris, iris-1}. We do not explain Iris in full detail here; we only focus on the features that are relevant to the thesis and omit or simplify some subtle details for presentation. For a comprehensive introduction, refer to the documents in the Iris webpage \cite{iris-note, ground-up, ralf-thesis}.

Throughout the thesis, we may skip some parts of the notation if the context is obvious.

\section{Separation Logic}

Concurrent separation logic \cite{csl} is a logic for concurrent programs. It is built around \textit{resources} that can be manipulated, composed, and split. Resources enable modular, thread-local reasoning: instead of reasoning on thread interleavings, we can reason within a thread with the resources it owns.

One of the most common resources is a \textit{points-to} assertion $\ell \mapsto v$, meaning that the heap has a location $\ell$ pointing to the value $v$. A thread owning $\ell \mapsto v$ is allowed to modify the value that $\ell$ points to.

For resources $P$ and $Q$, $P \ast Q$ is a resource called the \textit{separating conjunction}, which asserts that the heap can be split into two fragments, one satisfying $P$ and the other satisfying $Q$. Since the two parts must be disjoint, owning $\ell_1 \mapsto v_1 \ast \ell_2 \mapsto v_2$ implies $\ell_1 \neq \ell_2$, systematically preventing multiple threads from writing to the same location in an unwanted way. Being able to separate the parts of the heap enables local reasoning when desired, hiding away the parts that are not necessary for the proof target.

Next, $P \wand Q$ is a resource called the \textit{separating implication} or \text{magic wand}. It is a resource that, when combined with a heap satisfying $P$, asserts $Q$. In other words, we can combine $P$ and $P \wand Q$ to obtain $Q$, consuming the two in the process.

\begin{figure}
	\begin{mathpar}
		\axiomH{Points-To-Agree} {
\ell \mapsto^{q_1} v_1 \ast \ell \mapsto^{q_2} v_2 \wand v_1 = v_2
		}
		\and
		\axiomH{Points-To-Fractional} {
\ell \mapsto^{q_1+q_2} v \wandIff
\ell \mapsto^{q_1} v \ast \ell \mapsto^{q_2} v
(0 < q_1, 0 < q_2)
		}
		\\
		\axiomH{Load}
{\hoare{\ell \mapsto^q v}{! \ell}{u. u = v \ast \ell \mapsto^q v}}
		\and
		\axiomH{Store}
{\hoare{\ell \mapsto v}{\ell \la w}{\ell \mapsto w}}
	\end{mathpar}
	\caption{Proof rules for points-to.}
	\label{fig:pointsto}
\end{figure}

To share a location between threads, we use \textit{fractional points-to}: for a fraction $q$ where $0 < q \leq 1$, the resource $\ell \mapsto^q v$ represents a fractional ownership of the location $\ell$. A thread with any fractional ownership can read a value from it, but only a thread with full ($q = 1$) ownership can write a value to it. This reflects that there should be either one writer and no readers, or multiple readers and no writer, to avoid data race. Note that $\ell \mapsto^{1} v$ is the same as $\ell \mapsto v$. Fractional points-to can be split and distributed to threads, or vice-versa, and two points-to from the same location has the same value. These rules are presented formally in \autoref{fig:pointsto}.

\section{Specification and Invariant}

\label{section:hoare}
The specification of a program is represented with \textit{Hoare triples} of the form $\hoare{P}{e}{v. Q}$. This means that given a resource $P$, the computation of $e$ does not get stuck, and upon completion, it transforms $P$ to a resource $Q$ and returns a value $v$. These $P$ and $Q$ are called the \textit{precondition} and the \textit{postcondition}, respectively. We may skip $v$ if $e$ returns nothing.

For example, the bottom two rules of \autoref{fig:pointsto} shows two specifications for reading and writing to a location. \ref{Load} states that we can use any fractional points-to to read from it, and \ref{Store} states that we can use a full points-to to write to it.

The power of Hoare triples in separation logic is that we can use the following \textit{frame rule} to apply the specification for any heap containing $P$:

$$\infer { \hoare{P \ast R}{e}{Q \ast R} }{ {\hoare{P}{e}{Q}} } $$

Therefore, we can apply \ref{Load} or \ref{Store} in the presence of other resources and they will not be affected by it.

For a data structure, a Hoare triple usually doesn't hold by itself, because the correctness relies on the internal property being held by the data structure. An \textit{invariant} denoted $\knowInv{}{I}$ is used to express this property and to prove the specification: it says that the property $I$ holds at every program step.

Invariants are used with the following proof rule:

$$\infer
{\knowInv{}{I} \vdash \hoare{P}{e}{v. Q}}
{\hoare{I \ast P}{e}{v. I \ast Q} \quad \atomic(e)}
$$

This states that we can use the content $I$ of the invariant during an atomic instruction, and must give it back to the invariant after the instruction. $e$ being atomic ensures that the invariant is indeed satisfied at every step.

Using a Hoare triple and an invariant, the specification of \verb|push| would look like the following:

$$\knowInv{}{\dequeInv(p)} \vdash \hoare{\deque(p, \ell)}{push(p, v)}{\deque(p, \ell + [v])}$$

Unfortunately, this specification is not actually strong enough to be usable. We defer the discussion on a stronger specification to \autoref{section:lat}.

\section{Persistent Propositions}

Some resources are \textit{persistent}: once they hold, they stay true forever. Examples of persistent propositions are:
\begin{itemize}
	\item Invariants $\knowInv{}{I}$.
	\item Pure propositions such as equality and inequality.
	\item Persistent points-to $\ell \mapsto ^{\always} v$. This means $\ell$ is always a pointer pointing to the value $v$. Any fractional points-to can be made persistent, and once made so, it cannot be turned back to non-persistent and the corresponding value cannot be changed anymore.
\end{itemize}

Persistent resources cannot be transformed, but they have other useful properties. First, they can be freely duplicated ($P \wand P \ast P$) and shared among threads. In particular, if an invariant contains a persistent resource, we can keep it after closing the invariant by duplicating it. Also, if at least one side of the separating implication $P \wand Q$ is persistent, then $P$ does not disappear after using it to produce $Q$.

\section{Ghost States}

Iris supports \textit{ghost states}, purely logical resources that are not directly affected by program executions. Unlike physical resources like points-to, users verifying a program have freedom over the choice of ghost states and can manipulate them as they wish. By tying them with physical states via an invariant, they can ensure some properties that physical states themselves cannot express.

A ghost state is an element defined in a set of values called a \text{resource algebra} (RA), equipped with suitable operators satisfying certain conditions like commutativity and associativity (exact conditions are omitted here). Users may use multiple independent ghost states using \textit{ghost names}: the ghost states under different ghost names do not interfere with each other and may even be defined under different RAs. Given an element $a$ of an RA, and a ghost name $\gamma$, the proposition $\ownGhost{\gamma}{a}$ asserts ownership of the ghost state $a$ under the ghost name $\gamma$.

\begin{figure}
	\begin{mathpar}
		\axiomH{Ghost-Var-Alloc}
		{\forall a, \TRUE \vsWand \exists \gamma. \ghostVar{\gamma}{1}{a}}
		\\
		\axiomH{Ghost-Var-Valid}
		{\ghostVar{\gamma}{q}{a} \vdash 0 < q \leq 1}
		\and
		\axiomH{Ghost-Var-Agree}
		{\ghostVar{\gamma}{q_1}{a_1} \ast \ghostVar{\gamma}{q_2}{a_2} \vdash a_1 = a_2}
		\\
		\axiomH{Ghost-Var-Fractional}
		{\infer {
			\ghostVar{\gamma}{q_1 + q_2}{a} \wandIff
			\ghostVar{\gamma}{q_1}{a} \ast \ghostVar{\gamma}{q_2}{a}
			} {0 < q_1 \wedge 0 < q_2} }
		\and
		\axiomH{Ghost-Var-Update}
		{\ghostVar{\gamma}{1}{a} \vsWand \ghostVar{\gamma}{1}{b}}
	\end{mathpar}
	\caption{Proof rules for ghost variables.}
	\label{fig:ghostvar}
\end{figure}

Iris comes with many pre-defined ghost states. One of them is a \textit{ghost variable}, which is a resource similar to a fractional points-to but for a ghost (imaginary) location identified by $\gamma$. For ghost variables only, we will denote them as $\ghostVar{\gamma}{q}{a}$ instead of putting $\gamma$ outside of the dotted box. Ghost variables provide the rules in \autoref{fig:ghostvar}. Most rules are analogous to the fractional points-to rules, but \autoref{Ghost-Var-Update} uses a \textit{view shift} $\vsWand$. Here, $P \vsWand Q$ means we can transform $P$ to $Q$ by updating ghost states. Note that this is different from $\wand$: unlike a magic wand, a view shift may change the ghost states we own.

\begin{figure}
	\begin{mathpar}
		\axiomH{Mono-Nat-Alloc}
		{\forall n, \TRUE \vsWand \exists \gamma. \natauth{\gamma}{n}}
		\and
		\axiomH{Mono-Nat-Lb-Persistent}
		{\persistent{\natfrag{\gamma}{n}}}
		\\
		\axiomH{Mono-Nat-Auth-Exclusive}
		{\natauth{\gamma}{n} \ast \natauth{\gamma}{n'} \vdash \FALSE}
		\and
		\axiomH{Mono-Nat-Auth-Update}
		{\infer {\natauth{\gamma}{n} \vsWand \natauth{\gamma}{n'}} {n \leq n'}}
		\\
		\axiomH{Mono-Nat-Lb-Valid}
		{\natauth{\gamma}{n} \ast \natfrag{\gamma}{m} \vdash m \leq n}
		\and
		\axiomH{Mono-Nat-Lb-Get}
		{\natauth{\gamma}{n} \wand \natfrag{\gamma}{n}}
	\end{mathpar}
	\caption{Proof rules for monotonic natural numbers.}
	\label{fig:mononat}
\end{figure}

A more interesting example is a \textit{monotonic natural number}, the ghost state of natural numbers which can only increase, with proof rules given in \autoref{fig:mononat}. This is one of the \textit{authoritative ghost states} which consist of an authoritative form $\authfull$ and a snapshot form $\authfrag$. The authoritative form represents the exclusive ownership of the full information about the ghost state. This exclusiveness is expressed in \ref{Mono-Nat-Auth-Exclusive}. Since it asserts the full ownership, we can update it via \ref{Mono-Nat-Auth-Update}. However, being monotonic, the contained number can only increase. On the other hand, the snapshot form represents the persistent knowledge of the ghost state at some past point of execution. As such, using \ref{Mono-Nat-Lb-Valid}, we can infer that the number held by the authoritative form is at least the number taken by the past snapshot. Finally, we can take a new snapshot of the authoritative form using \ref{Mono-Nat-Lb-Get}. Note that this rule is not a view shift: it rather says that the ownership of $\natauth{\gamma}{n}$ automatically derives the ownership of $\natfrag{\gamma}{n}$. Since $\natfrag{\gamma}{n}$ is persistent, we retain $\natauth{\gamma}{n}$ after applying this rule.

\section{Linearizability and Logical Atomicity}

\label{section:lat}
While Hoare triples are intuitive, they are limited to sequential programs: since \verb|push| is not a physically atomic instruction, it is not possible to open an invariant around it. Consequently, a thread trying to use this specification must exclusively own the resource $WSDeque(p, \ell)$ instead of sharing it via an invariant. This defeats the purpose of supporting concurrent accesses to the data structure. Furthermore, the specification given in \autoref{section:hoare} does not make sense with concurrency: the contents of the deque before and after pushing may be completely different because other threads can change it in the meantime.

Instead, the specification of concurrent data structures is commonly given as \textit{linearizability} \cite{linearizability}. A concurrent data structure is linearizable if its concurrent invocations can be reordered to sequential invocations with the same effect on the data structure. For example, suppose the owner pushes a value to an empty deque, but right before returning from the operation, a stealer steals the same value. Although the deque was accessed concurrently, this has the same effect as pushing and then stealing sequentially.

A common way to prove the linearizability is to identify the \textit{commit points} of each operation: the point in which the operation ``appears'' to take place atomically from a client's viewpoint. Functions with such commit points are called \textit{logically atomic}. For \verb|push|, reading the bottom index does not affect the data structure; neither does writing a value to the array since it cannot be accessed by clients yet. Only after incrementing the bottom index is the whole effect of the operation visible to other clients. Therefore, incrementing the bottom is the commit point of \verb|push|. Once we identify the commit points of all other operations, we can order any set of concurrent invocations by the order in which their commit points are reached, which gives a linearization order of these invocations.

The idea of commit points is encoded in \textit{logically atomic triples} (LATs) \cite{prophecy}. They are denoted as $\ahoare{P}{e}{v. Q}$, which means that there is a commit point in $e$ which transforms the resource $P$ to $Q$, and $e$ returns $v$ in the end. To distinguish with Hoare triples, $P$ and $Q$ are called the \textit{atomic precondition} and the \textit{atomic postcondition}, respectively.

Just like physically atomic instructions, we can open an invariant around a logically atomic instruction:

$$\infer
{\knowInv{}{I} \vdash \ahoare{P}{e}{v. Q}}
{\ahoare{I \ast P}{e}{v. I \ast Q}}
$$

Of course, using this rule requires proving the LAT $\ahoare{I \ast P}{e}{v. I \ast Q}$, i.e. that $e$ really is logically atomic. As we can't just open invariants forever, how can a data structure prove that its methods are logically atomic? That is done by the following rule:

$$\infer
{\ahoare{P}{e}{v. Q}}
{\forall \Phi. \hoare{\AU{P}{Q}{\Phi}}{e}{\Phi}}
$$

Here, we receive an \textit{atomic update} $\AU{P}{Q}{\Phi}$ as a precondition. It is a resource representing the right and obligation to commit. At the commit point of $e$, we open $\au$ just like an invariant and obtain $P$. Then in a single physical step, we must transform $P$ into $Q$. Upon transformation, $\au$ is consumed and we get the assertion $\Phi$. Since $\Phi$ is universally quantified and contained in the postcondition, the only way to prove this Hoare triple is by completing the obligation of $\au$. This can be roughly formalized as
$
\au \vsWand (
	P \ast (
		Q \vsWand \Phi
	)
)
$, except for the single physical step requirement.

%% file: verif_simple.tex
\label{chap:verif-simple}
Now we are almost ready to give a specification to Chase-Lev deque and verify it. Before that, we need to extend the concept of logically atomic triples in order to give a proper specification. We start this chapter with the extension of LATs to support \textit{private postconditions}. Then we present the formal verification of Chase-Lev deque but without array resizing; it will be added back in the next chapter.

\section{Specification with Private Postconditions}

Using a logically atomic triple we saw in \autoref{chap:iris}, the modified specification of \verb|push| would look like the following:
$$\knowInv{}{\dequeInv(p)} \vdash \ahoare{\deque(p, l)}{push(p, v)}{\deque(p, l + [v])}$$

However, this specification is still incorrect: this allows concurrent pushes to the same deque $p$, which lead to data race on the bottom index or the array. Since Iris is a sound logic, such a specification cannot be proven either. The problem is that \verb|push| is not supposed to be called by multiple threads concurrently, but only by the sole owner. At the same time, we do have to use LATs because we want to prove linearizability.

What is missing is an exclusive resource for the owner of the deque $p$, which we will denote as $\ownDeque(p)$. Threads accessing a shared work-stealing deque have uneven authority over it. Specifically, only the owner of the deque may push or pop an element from it, and has a total control over some information that only \verb|push| and \verb|pop| may alter. $\ownDeque(p)$ represents this owner-exclusive permission to push or pop from the deque $p$.

Then is the following the correct specification for \verb|push|?
$$\knowInv{}{\dequeInv(p)} \vdash \ahoare{\deque(p, l) \ast \ownDeque(p)}{push(p, v)}{\deque(p, l + [v]) \ast \ownDeque(p)}$$

The answer is no: this \textit{still} allows concurrent pushes to $p$. Clients using the deque $p$ would share the permission to use $p$ by storing $\exists l. \deque(p, l)$ in the invariant. But with the specification like this, they can just store $\ownDeque(p)$ in the invariant as well and call \verb|push| concurrently. This is again an unsafe specification which cannot be proven either.

The root of the problem is that $\ownDeque$ is supposed to be stay local to the owner thread, not shared with an invariant. To represent this restriction, we move $\ownDeque$ outside of the LAT:
$$\knowInv{}{\dequeInv(p)} \vdash \ownDeque(p) \wand \ahoare{\deque(p, l)}{push(p, v)}{\deque(p, l + [v]) \ast \ownDeque(p)}$$

This resolves the problem for the atomic precondition, but the atomic postcondition still has an issue: $\ownDeque$ is consumed at the commit point so it cannot be used later. This may not be a problem for \verb|push| since the commit point is at the end of the operation anyway, but that is not the case for \verb|pop|.

We cannot move $\ownDeque$ out of the atomic postcondition, as that just separates $\ownDeque$ from the specification. We have to require this $\ownDeque$ as the postcondition, but at the end of the program, not at the commit point. This is called the \textit{private postcondition}. A LAT with private postcondition $R$ is denoted as $\ahoare{P}{e}{v.Q; R}$. When proving this LAT, we are given $\au_{P,Q}(R \wand \Phi)$. That is, upon committing, we receive $R \wand \Phi$; we can then normally progress through the proof, but at the end of the program we have to prove and consume $R$ in order to obtain $\Phi$ and finish the proof. Currently Iris' Coq formalization of LAT does not support private postconditions, so we slightly extended the framework for them.

\begin{figure}
	\begin{mathpar}
		\axiomH{New-Spec} {
\hoare{0 < n}{new\_deque(n)}{ p. \exists \gamma.
	\knowInv{}{\dequeInvg(p)} \ast \dequeg( []) \ast \ownDequeg(p)
} }
		\\
		\axiomH{Push-Spec} {
\knowInv{}{\dequeInv(p)} \vdash \ownDeque(p) \wand
\ahoare{\deque(l)}{push(p, v)}{\deque(l + [v]); \ownDeque(p)}
}
		\\
		\axiomH{Pop-Spec} { {
\begin{array}{l}
	\knowInv{}{\dequeInv(p)} \vdash \ownDeque(p) \wand
	\\
	\ahoare{\deque(l)}{pop(p)}{w. \exists l'.
		\deque(l') \ast
		\bigvee \left\{ \begin{array}{l}
			w = None \wedge l = l'
			\\
			\exists v. w = Some(v) \wedge l = l' + [v]
		\end{array} \right. 
		; \ownDeque(p)
	}
\end{array}
} }
		\\
		\axiomH{Steal-Spec} {
\knowInv{}{\dequeInv(p)} \vdash
\ahoare{\deque(l)}{steal(p)}{w. \exists l'.
	\deque(l') \ast
	\bigvee \left\{ \begin{array}{l}
		w = None \wedge l = l'
		\\
		\exists v. w = Some(v) \wedge l = [v] + l'
	\end{array} \right. 
} }
	\end{mathpar}
	\caption{Specification of Chase-Lev deque.}
	\label{fig:spec}
\end{figure}

Now we can finally give the specification of Chase-Lev deque as \autoref{fig:spec}. \ref{New-Spec} is just a regular Hoare triple because there is no concurrency involved here. \ref{Push-Spec} is straightforward. \ref{Pop-Spec} and \ref{Steal-Spec} have case analysis on whether the attempt succeeded or not. If successful, the top or bottom element is removed from the abstract state and the removed element is returned. Otherwise, the abstract state does not change and nothing is returned.

\section{Resource Definitions}

So far, we haven't defined the resources $\deque$, $\dequeInv$, and $\ownDeque$. The clients of the deque can just use our specification without delving into the details of each resource, but since we have to prove the specification, we should define each resource using the pre-defined propositions.

The definition of our deque resources use several ghost states, so we need multiple ghost names. Fortunately, we can group multiple ghost names into a single one\footnote{Formally, a ghost name is represented by a natural number, and multiple ghost names can be encoded into a single ghost name with a suitable bijection between $\mathbb{N}^k$ and $\mathbb{N}$.}. In our case, we define $\gamma$ as a tuple of three other ghost names $(\gamma_q, \gamma_{sw}, \gamma_{state})$.

We use one type of ghost state for each ghost name, for a total of three. For $\gamma_q$, we use a ghost variable representing the abstract state of the deque. This is shared between $\deque$ and $\dequeInv$. For $\gamma_{sw}$ (shorthand for ``single-writer''), we use a ghost variable for the values that only the owner of the deque can alter. This is shared between $\dequeInv$ and $\ownDeque$. Finally, for $\gamma_{state}$, we use a custom ghost state $\dqstauth$, encoding some properties about the synchronization guarantee in Chase-Lev deque.

$\deque$ is simply defined as a ghost variable for $\gamma_q$:

$$
\dequeg(l) :=
\exists \gamma_q.
\gamma = (\gamma_q, \_, \_) \ast \ghostVar{\gamma_q}{1/2}{l}
$$

$\dequeInv$ consists of the points-to for each field of the deque, and the three ghost states:

$$
\dequeInvg(p) :=
\begin{array}{l}
	\exists \gamma_q, \gamma_{sw}, \gamma_{state}, C, arr, top, bot, arr, L, t, b, pop.
	\\
	\bigsep \left\{
\begin{array}{l}
    \gamma = (\gamma_q, \gamma_{sw}, \gamma_{state})
    \wedge
    p = (C, top, bot)
    \\
    1 \leq t \leq b < t + |L|
	\\
	C \mapsto ^{\always} (arr, |L|)
	\\
    \ghostVar{\gamma_q}{1/2}{L[t..b)}
    \ast
    \ghostVar{\gamma_{sw}}{1/2}{(L, b, pop)}
    \\
    \dqstauthg{\gamma_{state}}{L, t, b} 
    \\
    arr \mapsto^{1/2} L
    \ast
    top \mapsto t
    \\
    bot \mapsto^{1/2} \begin{cases} b-1 & \text{if } pop = true \\ b & \text{otherwise} \end{cases}
\end{array}
	\right.
\end{array}
$$

Finally, $\ownDeque$ consists of the remaining points-to, and the ghost state for $\gamma_{sw}$:

$$
\ownDequeg(p) :=
\begin{array}{l}
  \exists \gamma_q, \gamma_{sw}, \gamma_{state}, \gamma_{era}, C, top, bot, arr, L, b.
  \\
  \bigsep \left\{
    \begin{array}{l}
    \gamma = (\gamma_q, \gamma_{sw}, \gamma_{state})
    \wedge
    p = (C, top, bot)
    \\
	C \mapsto ^{\always} (arr, |L|)
	\\
    \ghostVar{\gamma_{sw}}{1/2}{(L, b, \FALSE)}
    \\
    arr \mapsto^{1/2} L \ast bot \mapsto^{1/2} b
    \end{array}
  \right.
\end{array}
$$

In \autoref{chap:cld}, we listed several properties about Chase-Lev deque:
(1) only the owner can modify the bottom and the array;
(2) the top can only increase;
(3) once $t < b$ holds, it stays true and the top element is preserved until $t$ increases.
Among those, (1) is encoded in our invariant as fractional points-to. Since \verb|steal| does not have access to $\ownDeque$, they only get a half points-to for $arr$ and $bot$, so they cannot write to them. On the other hand, the owner of the deque can combine $\ownDeque$ with $\dequeInv$ to get a full points-to for them and use it to write a new value.

(2) and (3) are encoded as our custom ghost state we will call the \textit{deque state}. It is an authoritative ghost state, and there are two types of ghost states:
\begin{itemize}
	\item $\dqstauthg{\gamma}{L, t, b}$, the authoritative form, represents the ownership of the current state of the deque. The state consists of the whole array $L$, top index $t$, and bottom index $b$.
	\item $\dqstfragg{\gamma}{L, t, b}$, the snapshot form, represents the persistent knowledge of a past state of the deque.
\end{itemize}

\begin{figure}
	\begin{mathpar}
		\axiomH{Dqst-Auth-Alloc} {
|L| \neq 0 \vsWand \exists \gamma. \dqstauthg{\gamma}{L, 1, 1}
}
		\and
		\axiomH{Dqst-Frag-Persistent} {
\persistent{\dqstfragg{}{L, t, b}}		
}
		\\
		\axiomH{Dqst-Frag-Get} {
\dqstauthg{}{L, t, b} \wand \dqstfragg{}{L, t, b}
}
		\and
		\axiomH{Dqst-Frag-Valid} {
\dqstfragg{}{L_1, t_1, b_1} \ast \dqstauthg{}{L_2, t_2, b_2} \wand
t_1 \leq t_2 \wedge \preservation{1}{2}
}
		\\
		\axiomH{Dqst-Write-Array} {
\dqstauthg{}{L, t, b} \vsWand \dqstauthg{}{L[b \leftarrow v], t, b}
}
		\and
		\axiomH{Dqst-Push} {
b+1 < t + |L| \implies
\dqstauthg{}{L, t, b} \vsWand \dqstauthg{}{L, t, b+1}
}
		\\
		\axiomH{Dqst-Pop} {
t < b-1 \implies
\dqstauthg{}{L, t, b} \vsWand \dqstauthg{}{L, t, b-1}
}
		\and
		\axiomH{Dqst-CAS-Top} {
t < b \implies
\dqstauthg{}{L, t, b} \vsWand \dqstauthg{}{L, t+1, b}
}
	\end{mathpar}
	\caption{Proof rules of deque state.}
	\label{fig:deque_state_rule}
\end{figure}

The proof rules for deque state are listed in \autoref{fig:deque_state_rule}. \ref{Dqst-Auth-Alloc}, \ref{Dqst-Frag-Persistent}, and \ref{Dqst-Frag-Get} are similar to the proof rules for monotonic natural number as seen in \autoref{chap:iris}. \ref{Dqst-Write-Array}, \ref{Dqst-Push}, \ref{Dqst-Pop}, and \ref{Dqst-CAS-Top} are the update rules that change the internal states.\footnote{It turns out \ref{Dqst-Write-Array} and \ref{Dqst-Pop} don't have to be a view shift; a wand is sufficient. We used a view shift here for consistency, and either of them works in verifying the deque operations.}

\ref{Dqst-Frag-Valid} is the key rule that contains our reasoning of top element preservation. Here, $\preservation{1}{2}$ is a persistent proposition and the shorthand for $(t_1 = t_2 \wedge t_1 < b_1) \implies t_2 < b_2 \wedge L_1[t_1] = L_2[t_2]$. Put together, this rule means: given a snapshot of a past state 1, and the ownership of the current state 2, we learn $t_1 \leq t_2$; and if $t$ has not increased and $t < b$ initially held, then it still holds in the current state and the top element also stays the same.

\section{Verification of Each Function}

\subsection{New}

\begin{figure}
	\begin{mathpar}
		\axiomH{Alloc} {
\hoare{\TRUE}{AllocN(n, v)}{
	l. l \mapsto [v, v, \cdots, v] \wedge |l| = n
} }
		\\
		\axiomH{Ref} {
\hoare{\TRUE}{ref(v)}{
	l. l \mapsto v
} }
	\end{mathpar}
	\caption{Specification of allocation in HeapLang.}
	\label{fig:spec_alloc}
\end{figure}

To verify \verb|new_deque| (\autoref{fig:new}), we have to allocate all physical and ghost resources required by the specification. HeapLang has physical allocation rules given in \autoref{fig:spec_alloc}. Following the code, we allocate the array using \ref{Alloc}, and allocate the pair $(array, sz)$, top index, and bottom index using \ref{Ref}. Then we make the pair points-to persistent, and we have $C \mapsto ^{\always} (arr, n) \ast arr \mapsto L \ast top \mapsto 1 \ast bot \mapsto 1$, where $L = [v, v, \cdots, v]$ with length $n$.

Next, we allocate the ghost variables $\ghostVar{\gamma_q}{1}{[]}$ and $\ghostVar{\gamma_{sw}}{1}{(L, 1, \FALSE)}$ via \ref{Ghost-Var-Alloc}, and $\dqstauthg{\gamma_{state}}{L, t, b}$ via \ref{Dqst-Auth-Alloc}. Finally, we split the points-to and ghost variables, group the resources as $\dequeInv(p) \ast \deque(l) \ast \ownDeque(p)$, and turn $\dequeInv(p)$ as the invariant $\knowInv{}{\dequeInv(p)}$, which proves the postcondition.

\subsection{Push}

As we are skipping the resizing part for now, the implementation of \verb|push| should be changed as well. We use the same code as \autoref{fig:push}, except:
\begin{itemize}
	\item If $t + sz \leq b + 1$, we call \verb|diverge| which just runs an infinite loop. Since a Hoare triple only concerns with the condition at termination (remind that LATs are also proved using a Hoare triple with the atomic update $\au$), we do not need to consider the branch ending up with an infinite loop. Specifically, \verb|diverge| has the specification $\hoare{\TRUE}{diverge()}{\FALSE}$, and using this specification we obtain $\FALSE$ as a resource, which can prove anything.
	\item We skip \verb|"circle'" := !(arr "deque")| and \verb|"sz'" := Snd "circle"|, since there is no resizing. All other occurrences of \verb|"circle'"| and \verb|"sz'"| are replaced with \verb|"circle"| and \verb|"sz"| respectively.
\end{itemize}

Now we verify the variant of \verb|push| without resizing. In the beginning, we have $\ownDeque(p)$: $C \mapsto ^{\always} (arr, |L|) \ast \ghostVar{\gamma_{sw}}{1/2}{(L, b, \FALSE)} \ast arr \mapsto^{1/2} L \ast bot \mapsto^{1/2} b$. To read from the array or bottom, we just use the resources in $\ownDeque$ since we can use any fraction to read. To read from the top, we open the invariant to get $top \mapsto t$, read the value $t$ using it, and close the invariant.

To write to the array, we can't use $arr \mapsto^{1/2} L$; we should open the invariant to get the other half. Note that $\dequeInv$ has an existential for $L$, and opening the invariant does not guarantee by itself that $L$ from $\dequeInv$ equals $L$ from $\ownDeque$. That is where fractional resources come into play: since we have $arr \mapsto^{1/2} L'$ and $arr \mapsto^{1/2} L$, we can use \ref{Points-To-Agree} to prove $L' = L$.

Now we can combine them into $arr \mapsto L$, use \ref{Store} to write a new value, which turns it into $arr \mapsto L_2$ for some new list $L_2$, and split them back into $arr \mapsto^{1/2} L_2 \ast arr \mapsto^{1/2} L_2$. However, we cannot close the invariant yet. We would like to pick $L_2$ as its existential for $L$, but this does not work because we don't have $\ghostVar{\gamma_{sw}}{1/2}{(L_2, b', pop)}$. Instead we have $\ghostVar{\gamma_{sw}}{1/2}{(L, b', pop)}$, so we should update this.

To do so, we apply the same procedure as above but for ghost variables: since we have the other half $\ghostVar{\gamma_{sw}}{1/2}{(L, b, \FALSE)}$ from $\ownDeque$, we can use \ref{Ghost-Var-Agree} to prove $(L, b', pop) = (L, b, \FALSE)$. Then we combine the two identical ghost variables into $\ghostVar{\gamma_{sw}}{1}{(L, b, \FALSE)}$ by \ref{Ghost-Var-Fractional}, update it to $\ghostVar{\gamma_{sw}}{1}{(L_2, b, \FALSE)}$ by \ref{Ghost-Var-Update}, and split it back by \ref{Ghost-Var-Fractional}.

We also have to update $\dqstauthg{}{L, t, b}$ to $\dqstauthg{}{L_2, t, b}$, which can be done by \ref{Dqst-Write-Array}. We don't need to update $\ghostVar{\gamma_q}{1/2}{L[t..b)}$: we can instead prove $L_2[t..b) = L[t..b)$ because we actually wrote right next to the boundary of $[t..b)$, so there is nothing to update. Finally, we can prove $|L_2| = |L| = n$. Now that we changed all occurrences of $L$ to $L_2$, we can close the invariant by choosing $L_2$ as its existential for $L$.

After writing to the array, we have to increment the bottom. This is the commit point of \verb|push|. We open not only the invariant, but also the atomic update $\au$ to get the atomic precondition $\deque(l)$. Now we have to turn it into the atomic postcondition $\deque(l + [v])$, which can be done by combining the two ghost variables for $\gamma_q$ and updating them, just like what we did for writing to the array. After committing $\au$ and closing the invariant, we are left with $\ownDeque \wand \Phi$ and $\ownDeque$. We combine them to get $\Phi$, finishing the proof.

To describe the proof in more detail, let's progress through the program one by one:
\begin{itemize}
	\item Start with $\knowInv{}{\dequeInv} \ast \ownDeque$.
	\item \verb|"b" := !(bot "deque")|: read $b$ from $bot \mapsto^{1/2} b$.
	\item \verb|"t" := !(top "deque")|: we do not have a points-to for $top$, but it is hidden in $\dequeInv$. We proceed as follows:
	\\ (1) Open $\dequeInv$. We obtain all resources in $\dequeInv$ for some instantiation of its existentials, but this time we are only interested in $top \mapsto t$.
	\\ (2) Read $t$ from $top \mapsto t$.
	\\ (3) Close $\dequeInv$ with the same instantiation of its existentials by returning all resources for it including $top \mapsto t$.
	\item \verb|"circle" := !(arr "deque")|: read $C$ from $C \mapsto ^{\always} (arr, |L|)$.
	\item \verb|"sz" := Snd "circle"|: $sz = |L|$.
	\item Assume $t + sz > b + 1$, so there is a slot in the array to write $v$; otherwise we end up with an infinite loop so we are done.
	\item \verb|(access (Fst "circle") "b" "sz") <- "v"|: to write to the array, we need a full points-to. We proceed as follows:
	\\ (1) Open $\dequeInv$. Let $\gamma_q', \cdots, pop'$ be the instantiation for its existentials.
	\\ (2) From $\gamma = (\gamma_q, \gamma_{sw}, \gamma_{state})$ and $\gamma = (\gamma_q', \gamma_{sw}', \gamma_{state}')$, prove $\gamma_q = \gamma_q'$, $\gamma_{sw} = \gamma_{sw}'$, and $\gamma_{state} = \gamma_{state}'$. Similarly, prove $C = C'$, $top = top'$, and $bot = bot'$, and then prove $arr = arr'$ and $|L| = n'$.
	\\ (3) Use \ref{Ghost-Var-Agree} on $\gamma_{sw}$ to also prove $(L', b', pop') = (L, b, pop)$.
	\\ (4) Combine $arr \mapsto^{1/2} L \ast arr \mapsto^{1/2} L$ into $arr \mapsto L$, apply \ref{Store} to change it to $arr \mapsto L[b \leftarrow v]$, and split it back.
	\\ (5) Similarly, use \ref{Ghost-Var-Fractional} and \ref{Ghost-Var-Update} to update the ghost variable for $\gamma_{sw}$ to $\ghostVar{\gamma_{sw}}{1/2}{(L[b \leftarrow v], b, pop)}$.
	\\ (6) Update $\dqstauthg{}{L, t, b}$ to $\dqstauthg{}{L[b \leftarrow v], t, b}$ by \ref{Dqst-Write-Array}.
	\\ (7) Close $\dequeInv$ with $L[b \leftarrow v]$ as the choice of $L$.
	\item \verb|bot "deque" <- "b" + 1|: this is the commit point.
	\\ (1) Open $\dequeInv$ and prove $(L', b', pop) = (L[b \leftarrow v], b, \FALSE)$ by \ref{Ghost-Var-Agree}.
	\\ (2) Combine and update the resources to $bot \mapsto b+1$ by \ref{Store}, $\ghostVar{\gamma_{sw}}{1}{(L[b \leftarrow v], b+1, \FALSE)}$ by \ref{Ghost-Var-Update}, and $\dqstauthg{}{L[b \leftarrow v], t, b+1}$ by \ref{Dqst-Push}, then split them back as required.
	\\ (3) Open the atomic update $\au$ which contains $\deque$ and prove $l = L[b \leftarrow v][t..b)$ by \ref{Ghost-Var-Agree}. This also implies $l + [v] = L[b \leftarrow v][t..b+1)$.
	\\ (4) Combine and update the resources to $\ghostVar{\gamma_q}{1}{l + [v]}$ by \ref{Ghost-Var-Update}, and split it back.
	\\ (5) Now we have the atomic postcondition $\deque(l + [v])$, so commit $\au$ and obtain $\Phi$.
	\\ (6) Close the invariant with $b+1$ as the choice of $b$.
	\item In the end, we have $\ownDeque(p) \ast (\ownDeque \wand \Phi)$. Combine them into $\Phi$ and finish the proof.
\end{itemize}

\subsection{Pop}

Verification of \verb|pop| roughly follows a similar procedure to verifying \verb|push|, except for some key differences. First, instead of changing $L$, we change $pop$. This value is initially $\FALSE$. When we decrement the bottom in the beginning, we set it to $\TRUE$ and update the ghost variables accordingly. When we increment it back, or end up in the normal case, we set it back to $\FALSE$ and update the ghost variables again.

Another difference is that the commit point depends on the number of elements in the deque when reading the top.
If the deque is empty, reading the top is the commit point since that's when we observe the empty deque. The atomic precondition and postcondition have the same resources in this case, so we can just open $\au$ and commit immediately.
If the deque has more than one element, reading the top is still the commit point.
If the deque has exactly one element, CAS-ing the top is the commit point, regardless of whether the CAS succeeds or not.

Note that this is not the only possible way to determine the commit points. For example, we believe that the following approach also works: if the deque has more than one element, decrementing the bottom is the commit point. In fact, this may intuitively make more sense to some because that's when the bottom element is removed. This approach, however, has a big downside: we have to do a case analysis on which value of the top index \textit{will be read in the next step}. This reasoning is possible and actually necessary in some data structures \cite{prophecy}, but it's complicated and beyond the scope of this thesis.

Here is the proof outline in more detail:

\begin{itemize}
	\item Start with $\knowInv{}{\dequeInv} \ast \ownDeque$.
	\item \verb|"b" := !(bot "deque") - #1|: read $b$ from $bot \mapsto^{1/2} b$.
	\item \verb|"circle" := !(arr "deque")|: read $(arr, |L|)$ from $C \mapsto ^{\always} (arr, |L|)$.
	\item \verb|"sz" := Snd "circle"|: $sz = |L|$.
	\item \verb|bot "deque" <- "b"|: open $\dequeInv$, update $bot \mapsto b$ to $bot \mapsto b-1$, update $\ghostVar{\gamma_{sw}}{1}{(L, b, \FALSE)}$ to $\ghostVar{\gamma_{sw}}{1}{(L, b, \TRUE)}$, and close $\dequeInv$ with $\TRUE$ as the choice of $pop$.
	\item \verb|"t" := !(top "deque")|:
	\\ (1) Open $\dequeInv$ and read $t$ from $top \mapsto t$.
	\\ (2) If $t < b$ (normal case), open $\au$ as well, update the ghost states to $\ghostVar{\gamma_q}{1}{L[t..b-1)} \ast \ghostVar{\gamma_{sw}}{1}{(L, b-1, \FALSE)} \ast \dqstauthg{}{L, t, b-1}$ by \ref{Dqst-Pop}, and commit.
	\\ (3) If $b < t$ (empty case), open $\au$ and immediately commit.
	\\ (4) Otherwise, do nothing. In all cases, we close $\dequeInv$ in the end.
	\item If $b < t$ (empty case), \verb|bot "deque" <- "t"|: open $\dequeInv$, update $bot \mapsto b-1$ to $bot \mapsto b$, update $\ghostVar{\gamma_{sw}}{1}{(L, b, \TRUE)}$ back to $\ghostVar{\gamma_{sw}}{1}{(L, b, \FALSE)}$, and close $\dequeInv$ with $\FALSE$ as the choice of $pop$. We already committed, so we have $\ownDeque \ast \Phi$ and we are done.
	\item \verb|"v" := !(access (Fst "circle") "b" "sz")|: read $v$ from $arr \mapsto^{1/2} L$.
	\item If $t < b$ (normal case), we already committed so we are done.
	\item \verb|CAS (top "deque") "t" ("t" + #1)|: open $\dequeInv$ and attempt CAS with $top \mapsto t'$.
	\item If $t = t'$, we succeed the CAS and get $top \mapsto t+1$. Open $\au$, update the ghost states to $\ghostVar{\gamma_q}{1}{L[t+1..b)} \ast \dqstauthg{}{L, t+1, b}$, commit, and close $\dequeInv$ with $t+1$ as the choice of $t$.
	\item If $t \neq t'$, we fail the CAS. Open $\au$, immediately commit, and close $\dequeInv$.
	\item \verb|bot "deque" <- "t" + #1|: open $\dequeInv$, update $bot \mapsto b-1$ to $bot \mapsto b$, update $\ghostVar{\gamma_{sw}}{1}{(L, b, \TRUE)}$ to $\ghostVar{\gamma_{sw}}{1}{(L, b, \FALSE)}$, and close $\dequeInv$ with $\FALSE$ as the choice of $pop$.
	\item In the end, we have $\Phi \ast \ownDeque$ so we are done.
\end{itemize}

\subsection{Steal}

When verifying \verb|steal|, we don't have $\ownDeque$, so we cannot keep track of the array and bottom. Let $t_i$, $b_i$, and $L_i$ be the top, bottom, and array, respectively, when $\dequeInv$ is opened for the $i$-th time. We open the invariant four times: (1) reading the top, (2) reading the bottom, (3) reading the value at the top, and (4) CAS-ing the top.

The commit points are similar to \verb|pop|. If the deque is ``empty'' (no element, or one element and the owner is trying to pop), reading the bottom is the commit point. Otherwise, CAS-ing the top is the commit point.

The main challenge in verifying \verb|steal| is to prove that the value at the top is preserved at the successful CAS. Specifically, we have to prove $L_3[t_3] = L_4[t_4]$. The reason is that when opening $\au$ to get $\deque(l)$ and committing, we should prove $\exists l'. l = [L_3[t_3]] + l'$ and update $\ghostVar{\gamma_q}{1}{l}$ to $\ghostVar{\gamma_q}{1}{l'}$. We have $l = L_4[t_4..b_4)$ from $\ghostVar{\gamma_q}{1/2}{l}$ and $\ghostVar{\gamma_q}{1/2}{L_4[t_4..b_4)}$, so the first element of $l$ is $L_4[t_4]$. Therefore our goal reduces to $L_3[t_3] = L_4[t_4]$.

The intuitive reason why this holds is because if CAS succeeded, the top has never increased between reading the top and CAS-ing it, i.e. $t_1 = t_2 = t_3 = t_4$. Since we did not take the ``no chance'' branch, we have $t_1 < b_2$, so $t_2 < b_2$ as well. We saw in \autoref{chap:cld} that once $t < b$ holds, $L[t]$ is preserved until $t$ increases; therefore, $L_3[t_3] = L_4[t_4]$.

This reasoning can be expressed in Iris using our deque state $\dqstauth$. Each time we open the invariant, we take the persistent snapshot $\dqstfrag$ by \ref{Dqst-Frag-Get}. From the second opening onwards, we use $\dqstauth$ from the current opening, $\dqstfrag$ from the previous opening, and \ref{Dqst-Frag-Valid} to prove that the top has not decreased. This gives $t_1 \leq t_2 \leq t_3 \leq t_4$. If CAS succeeded, we know $t_1 = t_4$, so all of the $t_i$ must equal. To prove $L_3[t_3] = L_4[t_4]$, during the third opening, we use $\dqstauth$, the second $\dqstfrag$, and \ref{Dqst-Frag-Valid} to get $\preservation{2}{3}$. Then during the fourth opening, we learn $t_2 = t_3$ and thus $t_3 < b_3$. Finally, we apply \ref{Dqst-Frag-Valid} again to get $\preservation{3}{4}$ and thus $L_3[t_3] = L_4[t_4]$.

Here is the proof outline in more detail:

\begin{itemize}
	\item We start with just $\knowInv{}{\dequeInv}$.
	\item \verb|"t" := !(top "deque")|: open $\dequeInv$, read $t_1$ with $top \mapsto t_1$, take a snapshot $\dqstfrag(L_1, t_1, b_1)$ by \ref{Dqst-Frag-Get}, duplicate $C \mapsto ^{\always} (arr, n)$ and keep it locally, and close $\dequeInv$.
	\item \verb|"b" := !(bot "deque")|: open $\dequeInv$, read $b_2$ with $bot \mapsto b_2$, use $\dqstauth$ to prove $t_1 \leq t_2$ by \ref{Dqst-Frag-Valid}, take a snapshot $\dqstfrag(L_2, t_2, b_2)$, and close $\dequeInv$. Note that $b_2$ is not necessarily the same as $b$ in $\dequeInv$'s existential, because $pop$ may be $\TRUE$. If $b_2 \leq t_1$, open $\au$ and immediately commit.
	\item \verb|"circle" := !(arr "deque")|: read $(arr, n)$ from $C \mapsto ^{\always} (arr, n)$.
	\item \verb|"sz" := Snd "circle|: $sz = n$.
	\item If $b_2 \leq t_1$ (no chance), we already committed and we are done.
	\item \verb|"v" := !(access (Fst "circle") "t" "sz")|: open $\dequeInv$, read $v = L_3[t_3]$ with $arr \mapsto^{1/2} L_3$, use $\dqstauth$ to prove $t_2 \leq t_3 \wedge \preservation{2}{3}$ by \ref{Dqst-Frag-Valid}, take a snapshot $\dqstfrag(L_3, t_3, b_3)$, and close $\dequeInv$.
	\item \verb|CAS (top "deque") "t" ("t" + #1)|: open $\dequeInv$, use $\dqstauth$ to prove $t_3 \leq t_4$ by \ref{Dqst-Frag-Valid}, and attempt CAS with $top \mapsto t_4$.
	\item If $t_1 = t_4$:
	\\ (1) We succeed the CAS and we get $top \mapsto t_1+1$. Since $t_1 \leq t_2 \leq t_3 \leq t_4 = t_1$, we get $t_1 = t_2 = t_3 = t_4$ and in particular $t_2 = t_1 < b_2$.
	\\ (2) Use $\preservation{2}{3}$ and $t_2 = t_3 \wedge t_2 < b_2$ to prove $t_3 < b_3$.
	\\ (3) Use $\dqstauth$ and $\dqstfrag(L_3, t_3, b_3)$ to prove $t_4 < b_4$ and $L_3[t_3] = L_4[t_4] = v$ by \ref{Dqst-Frag-Valid}.
	\\ (4) Open $\au$, update the ghost states to $\ghostVar{\gamma_q}{1}{L_4[t_4+1..b_4)} \ast \dqstauth{}{L_4, t_4+1, b_4}$ by \ref{Dqst-CAS-Top}, commit, and close $\dequeInv$.
	\item If $t_1 \neq t_4$, we fail the CAS. Open $\au$, immediately commit, and close $\dequeInv$.
	\item In the end in both cases, we have $\Phi$ which finishes the proof.
\end{itemize}

\section{Construction of the Ghost Deque State}

So far, we gave the specification of Chase-Lev deque, defined the resources necessary for the specification, and used them to verify each operation. But there is one last problem remaining: we have to prove that the ghost deque state $\dqstauth$ and $\dqstfrag$ actually make sense. The proof rules for Iris' built-in ghost states like ghost variables and monotonic natural numbers have already been proven sound, but not our custom ghost state.

\begin{figure}
	\begin{mathpar}
		\axiomH{Ghost-Map-Alloc} {
True \vsWand \exists \gamma. \mapauth{\gamma}{\emptyset}
}
		\and
		\axiomH{Ghost-Map-Elem-Persistent} {
\persistent{\mapfrag{\gamma}{k}{v}}
}
		\\
		\axiomH{Ghost-Map-Elem-Agree} {
\mapfrag{\gamma}{k}{v_1} \ast \mapfrag{\gamma}{k}{v_2} \wand v_1 = v_2
}
		\and
		\axiomH{Ghost-Map-Lookup} {
\mapauth{\gamma}{M} \ast \mapfrag{\gamma}{k}{v} \wand M[k] = v
}
		\\
		\axiomH{Ghost-Map-Insert} {
k \notin \dom(M) \implies \mapauth{\gamma}{M} \vsWand \mapauth{\gamma}{M[k \leftarrow v]} \ast \mapfrag{\gamma}{k}{v}
}
	\end{mathpar}
	\caption{Proof rules of ghost map.}
	\label{fig:ghost_map}
\end{figure}

Fortunately, we can simply define our ghost states using other pre-defined ghost states, just like the resources in the specification like $\deque$. To define the ghost deque state, we need one more type of built-in ghost states called a \textit{ghost map}. This ghost state maintains a finite partial map from a domain set to a value set. $\mapauth{\gamma}{M}$ represents the authoritative ownership of the whole map $M$. For each key $k$ with its associated value $v$, $\mapfrag{\gamma}{k}{v}$ represents the persistent knowledge of this key-value pair.\footnote{There is also a ghost state for non-persistent key-value ownership so that we can change its associated value, but we don't use it here.} We can insert a new key-value pair $k \mapsto v$ into $M$, provided that $k$ is not mapped in $M$ already. The proof rules of ghost map are listed in \autoref{fig:ghost_map}.

\begin{figure}
\begin{mathpar}

\dqstauthg{\gamma}{L, t, b} :=
\begin{array}{l}
  \exists \gamma_{tb}, \gamma_{elt}, M.
  \\
  \bigsep \left\{
    \begin{array}{l}
    \gamma = (\gamma_{tb}, \gamma_{elt})
    \\
    \natauth{\gamma_{tb}}{tbs(t, b)}
    \\
    \mapauth{\gamma_{elt}}{M} \ast
    (t = b \vee \mapfrag{\gamma_{elt}}{t}{L[t]})
    \\
	\forall k \in \dom(M). k \leq \begin{cases}t & \text{if }t < b \\ t-1 & \text{otherwise}\end{cases}
    \end{array}
  \right.
\end{array}

\dqstfragg{\gamma}{L, t, b} :=
\begin{array}{l}
  \exists \gamma_{tb}, \gamma_{elt}.
  \\
  \bigsep \left\{
    \begin{array}{l}
    \gamma = (\gamma_{tb}, \gamma_{elt})
    \\
    \natfrag{\gamma_{tb}}{tbs(t, b)}
    \\
    t = b \vee \mapfrag{\gamma_{elt}}{t}{L[t]}
    \end{array}
  \right.
\end{array}
\end{mathpar}

\caption{The definition of ghost deque state.}
\label{fig:deque_state_def}
\end{figure}

We are finally ready to define the deque state. Remind that deque state should encode two properties:
(1) the top only increases, and
(2) the inequality between the top and bottom, and the top element are preserved along with the top.

For (1), we simply use a monotonic natural number. But we can use the same ghost state to also represent one half of (2). Specifically, let $tbs(t, b) := \begin{cases}2t+1 & \text{if }t < b \\ 2t & \text{otherwise}\end{cases}$. Then we can prove:

$$
tbs(t_1, b_1) \leq tbs(t_2, b_2) \iff
t_1 \leq t_2 \wedge (t_1 = t_2 \wedge t_1 < b_1 \implies t_2 < b_2)
$$

Therefore, the ownership of $\natauth{\gamma}{tbs(t, b)}$ ensures that (1) $t$ only increases, and (2) once $t < b$ holds, it keeps true until $t$ increases.

For the other half of (2), we use a ghost map $\mapauth{\gamma}{M}$, assigning to each top index its preserved top element. Each mapping $\mapfrag{\gamma}{i}{v}$ represents that $v$ is the preserved top element when the top index was $i$. If $t = b$, there is a mapping for each $i$ from $1$ to $t-1$. If $t < b$, there is a mapping up to $t$. To maintain this information, we insert a new mapping $t \mapsto L[t]$ whenever the top is incremented and it's still smaller than the bottom, or the bottom is incremented when the top and bottom were equal.

The full definition of the ghost deque state is given in \autoref{fig:deque_state_def}. The proof rules in \autoref{fig:deque_state_rule} are proven as follows:
\begin{itemize}
	\item \ref{Dqst-Auth-Alloc}: allocate $\natauth{\gamma_{tb}}{2}$ by \ref{Mono-Nat-Alloc} and $\mapauth{\gamma_{elt}}{\emptyset}$ by \ref{Ghost-Map-Alloc}. The other propositions are held automatically.
	\item \ref{Dqst-Frag-Persistent}: follows from \ref{Mono-Nat-Lb-Persistent}, \ref{Ghost-Map-Elem-Persistent}, and the persistence of pure propositions.
	\item \ref{Dqst-Frag-Get}: use \ref{Mono-Nat-Lb-Get}.
	\item \ref{Dqst-Frag-Valid}: use \ref{Mono-Nat-Lb-Valid} to prove $t_1 \leq t_2 \wedge (t_1 = t_2 \wedge t_1 < b_1 \implies t_2 < b_2)$. Next, given $t_1 < b_1$ and $t_2 < b_2$, we have $\mapfrag{\gamma_{elt}}{t_1}{L_1[t_1]} \ast \mapfrag{\gamma_{elt}}{t_2}{L_2[t_2]}$, so \ref{Ghost-Map-Elem-Agree} proves $L_1[t_1] = L_2[t_2]$.
	\item \ref{Dqst-Write-Array}: If $t = b$, trivial. Otherwise, $L[t] = L[b \leftarrow v][t]$ since $t \not\equiv b (\text{mod } |L|)$.
	\item \ref{Dqst-Push}: update $\natauth{\gamma_{tb}}{tbs(t, b)}$ to $\natauth{\gamma_{tb}}{tbs(t, b+1)}$ by \ref{Mono-Nat-Auth-Update}. If $t = b$, also update $\mapauth{\gamma_{elt}}{M}$ to $\mapauth{\gamma_{elt}}{M[t \leftarrow L[t]]}$ by \ref{Ghost-Map-Insert} and obtain $\mapfrag{\gamma_{elt}}{t}{L[t]}$.
	\item \ref{Dqst-Pop}: trivial.
	\item \ref{Dqst-CAS-Top}: update $\natauth{\gamma_{tb}}{tbs(t, b)}$ to $\natauth{\gamma_{tb}}{tbs(t+1, b)}$ by \ref{Mono-Nat-Auth-Update}. If $t+1 < b$, also update $\mapauth{\gamma_{elt}}{M}$ to $\mapauth{\gamma_{elt}}{M[t+1 \leftarrow L[t+1]]}$ by \ref{Ghost-Map-Insert} and obtain $\mapfrag{\gamma_{elt}}{t+1}{L[t+1]}$.
\end{itemize}

%% file: verif.tex
\label{chap:verif}
In this chapter, we extend the verification in \autoref{chap:verif-simple} to the full version, where the array resizes on overflow. We give a specification for resizing the array, extend the invariant and other resource definitions to account for array replacement, and list the changes in the verification of each operation.

\section{Specification of Resizing}

Recall from \autoref{chap:cld} that \verb|grow| (\ref{fig:push}) takes a circular array $circle$, top $t$, and bottom $b$, and returns a new circular array $circle'$ such that $circle[t..b) = circle'[t..b)$. The implementation is skipped in the paper, but on the high-level it operates as follows: we first read the size $n$ of $circle$ (the size is given in $circle$ along with the array) and allocate a new circular array $circle'$ with size $2n$. Then for each index $i \in [t..b)$, we read the index $i$ in the array of $circle$, and copy the value to the index $i$ in the array of $circle'$. The new $circle'$ is then returned.

The correctness of \verb|grow| relies on the fact that the array is used read-only. If the contents of the array change while copying a range, the new array may not make much sense. This assumption indeed holds because only the sole owner of the deque calls this function. Therefore, all we need is a fractional points-to for the array. But since the owner keeps track a half points-to locally, we do not need to open any invariant around it. We can thus give the specification of \verb|grow| as the following regular Hoare triple:

$$
t \leq b < t + |L| \implies
\hoare
{arr \mapsto^{1/2} L}
{grow((arr, |L|), t, b)}
{(arr', |L'|).
  \begin{array}{l}
    |L| < |L'| \wedge L[t..b) = L'[t..b) \ast
    \\
    arr \mapsto^{1/2} L \ast arr' \mapsto L'
  \end{array}
}
$$

The proof is a straightforward induction on $b-t$, so we omit the details here.

\section{New Deque State and Invariant}

Since we are removing the assumption that the array is never replaced, the resource definitions should change as well. Specifically, $C \mapsto ^{\always} \cdots$ no longer works because $C$ can be overwritten by resizing the array. Therefore we change them to $C \mapsto^{1/2} \cdots$ for both $\dequeInv$ and $\ownDeque$.

If we attempt the verification with these modified resource definitions, we still encounter a problem at \verb|steal| because we don't keep track of the old arrays. When reading the array, we open the invariant to get $C \mapsto^{1/2} (arr, n)$, read $(arr, n)$ from it, and close the invariant. Now all resources about $arr$ are lost: there is no local resource to synchronize because stealers do not have $\ownDeque$. When opening the invariant again, we get a different array $arr'$ as its existential. As a result, we get $arr' \mapsto \cdots$ even though we have to read from $arr$.

To receive the points-to for $arr$ at that point, the invariant should maintain not only the current array but also all the arrays that have been used so far. Informally, divide the timeline of operations as \textit{eras}, where a new era starts whenever the array is replaced. Then we extend the ghost deque state to contain the \textit{archive} of the information about all the past arrays: the array at the moment of archival along with its length, the points-to to that array, and the value of $tbs(t, b)$ at the moment of archival. Each era is represented by a ghost name $\gamma_{era}$ since we eventually have to tie a ghost state with each era.

The information \textit{at the moment of archival} poses a problem if we use $\dqstfrag$, because it could mean any point of execution, not just the moment of archival. We cannot use $\dqstauth$ either because the stealer cannot access the $\dqstauth$ at the moment of archival; only the owner can. Instead, we introduce another form of deque state, representing the information precisely at archival, which can be accessed from $\dqstauth$ of a future era.

Now there are three forms of deque state: $\dqstauthf{\gamma}{\gamma_{era}, arr, L, t, b}$ and $\dqstfragf{\gamma}{\gamma_{era}, arr, L, t, b}$ representing the full ownership and persistent knowledge of a deque state respectively, and $\dqstarchf{\gamma}{\gamma_{era}, arr, L, t, b}$ representing the ownership of a deque state of a \textit{past era}. We will call it the \textit{archived} deque state. Since we are going to put a points-to to the array in $\dqstauth$ for each past era, these resources are not technically ghost states anymore; thus we will not draw dotted lines around it.

\begin{figure}
	\begin{mathpar}
		\axiomH{Dqst-Archived-Frag-Get} {
\dqstarchf{}{\gamma, arr, L, t, b} \wand \dqstfragf{}{\gamma, arr, L, t, b}
}
		\and
		\axiomH{Dqst-Archived-Frag-Valid} {
\dqstfragf{}{\gamma, arr, L_1, t_1, b_1} \ast \dqstarchf{}{\gamma, arr, L_2, t_2, b_2} \wand
t_1 \leq t_2 \wedge \preservation{1}{2}
}
		\\
		\axiomH{Dqst-Frag-Agree} {
\dqstfragf{}{\gamma, arr_1, L_1, t_1, b_1} \ast \dqstfragf{}{\gamma, arr_2, L_2, t_2, b_2} \wand
arr_1 = arr_2 \wedge |L_1| = |L_2|
}
		\\
		\axiomH{Dqst-Archived-Get} {
{ \begin{array}{l l}
	& \gamma_1 \neq \gamma_2
		\wedge \dqstfragf{}{\gamma_1, arr_1, L_1, t_1, b_1}
		\ast \dqstauthf{}{\gamma_2, arr_2, L_2, t_2, b_2}
	\\
	\wand & \exists L, t, b.
	\bigsep \left\{
		\begin{array}{l}
			\dqstarchf{}{\gamma_1, arr_1, L, t, b} \ast arr_1 \mapsto L \\
			\dqstarchf{}{\gamma_1, arr_1, L, t, b} \ast arr_1 \mapsto L
			\wand \dqstauthf{}{\gamma_1, arr_1, L_1, t_1, b_1}
		\end{array}
	\right.
\end{array} }
}
		\\
		\axiomH{Dqst-Archive} {
{ \begin{array}{l l}
	& |L| \leq |L'| \wedge L[t..b) = L'[t..b) \ast arr \mapsto L
	\\
	\wand & \dqstauthf{}{\gamma, arr, L, t, b}
		\vsWand \exists \gamma'. \dqstauthf{}{\gamma', arr', L', t, b}
\end{array} }
}
	\end{mathpar}
	\caption{Additional proof rules of deque state with eras.}
	\label{fig:deque_state_more_rule}
\end{figure}

All of the proof rules in \autoref{fig:deque_state_rule} still apply to our extended deque states, with some straightforward modifications to include the era and the circular array pointer. \ref{Dqst-Auth-Alloc} may be less trivial to modify: the first era, $\gamma_0$, is made from the deque state allocation, i.e. $\exists \gamma, \gamma_0. \dqstauthf{\gamma}{\gamma_0, arr, L, 1, 1}$.

In addition, we use more proof rules given in \autoref{fig:deque_state_more_rule}. \ref{Dqst-Archived-Frag-Get} and \ref{Dqst-Archived-Frag-Valid} are similar to \ref{Dqst-Frag-Get} and \ref{Dqst-Frag-Valid} but from $\dqstarch$. \ref{Dqst-Frag-Agree} shows that the array pointer and the length of the array does not change within an era. \ref{Dqst-Archived-Get} lets us access an $\dqstarch$ of a past era from the $\dqstauth$. This has a pattern of $P \wand (Q \ast Q \wand P)$: this means that we have an access to $Q$ ``inside'' $P$, and we can return this $Q$ to get $P$ back. Finally, \ref{Dqst-Archive} lets us replace the array and advance the era, as long as the circular slice of $[t..b)$ is preserved.

Now we change $\dequeInv$ and $\ownDeque$ as follows. Not much is different from the definition given in \autoref{chap:cld}. The differences are: (1) the ghost variable for $\gamma_{sw}$ now contains $\gamma_{era}$ and $arr$ as well; (2) $\dqstauth$ has additional parameters as defined above; and (3) $C \mapsto \cdots$ is no longer persistent but only fractional.

$$
\dequeInvg(p) :=
\begin{array}{l}
	\exists \gamma_q, \gamma_{sw}, \gamma_{state}, \gamma_{era}, C, arr, top, bot, arr, L, t, b, pop.
	\\
	\bigsep \left\{
\begin{array}{l}
    \gamma = (\gamma_q, \gamma_{sw}, \gamma_{state})
    \wedge
    p = (C, top, bot)
    \\
    1 \leq t \leq b < t + |L|
	\\
	C \mapsto ^{1/2} (arr, |L|)
	\\
    \ghostVar{\gamma_q}{1/2}{L[t..b)}
    \ast
    \ghostVar{\gamma_{sw}}{1/2}{(\gamma_{era}, arr, L, b, pop)}
    \\
    \dqstauthf{\gamma_{state}}{\gamma_{era}, arr, L, t, b}
    \\
    arr \mapsto^{1/2} L
    \ast
    top \mapsto t
    \\
    bot \mapsto^{1/2} \begin{cases} b-1 & \text{if } pop = true \\ b & \text{otherwise} \end{cases}
\end{array}
	\right.
\end{array}
$$

$$
\ownDequeg(p) :=
\begin{array}{l}
  \exists \gamma_q, \gamma_{sw}, \gamma_{state}, \gamma_{era}, C, top, bot, arr, L, b.
  \\
  \bigsep \left\{
    \begin{array}{l}
    \gamma = (\gamma_q, \gamma_{sw}, \gamma_{state})
    \wedge
    p = (C, top, bot)
    \\
	C \mapsto ^{1/2} (arr, |L|)
	\\
    \ghostVar{\gamma_{sw}}{1/2}{(\gamma_{era}, arr, L, b, \FALSE)}
    \\
    arr \mapsto^{1/2} L \ast bot \mapsto^{1/2} b
    \end{array}
  \right.
\end{array}
$$

\section{Changes in the Verification}

Now we verify each operation again. Verifying \verb|new_deque| has no interesting difference; it's still just a matter of allocating resources. Verifying \verb|pop| is not much different either; no resizing takes place here. The ones that need extra proof work are \verb|push| because we have to resize the array, and \verb|steal| because we have to consider the case where the array is replaced by the owner.

\subsection{Push}

The main difference in the verification of \verb|push| is the branch where $t + sz \leq b + 1$, in which case we resize the array. We proceed as follows:
\begin{itemize}
	\item \verb|grow_circle "circle" "t" "b"|: Apply the specification of \verb|grow| with $arr \mapsto^{1/2} L$ to step through it and obtain a new points-to $arr_{new} \mapsto L_{new}$ such that $L[t..b) = L_{new}[t..b)$.
	\item \verb|arr "deque" <- ...|: This is the part where we resize the array.
	\\ (1) Open $\dequeInv$ and use \ref{Ghost-Var-Agree} to prove $(\gamma_{era}', arr', L', b', pop') = (\gamma_{era}, arr, L, b, \FALSE)$.
	\\ (2) Combine two halves of $arr \mapsto^{1/2} L$ into $arr \mapsto L$, then use \ref{Dqst-Archive} to consume it and update $\dqstauthf{\gamma_{dqst}}{\gamma_{era}, arr, L, t, b}$ to $\dqstauthf{\gamma_{dqst}}{\gamma_{era-new}, arr_{new}, L_{new}, t, b}$.
	\\ (3) Update the ghost variable for $\gamma_{sw}$ to $\ghostVar{\gamma_{sw}}{1/2}{(\gamma_{era-new}, arr_{new}, L_{new}, b, \FALSE)}$.
	\\ (4) Use \ref{Dqst-Frag-Get} to get $\dqstfrag$.
	\\ (5) Combine $C \mapsto^{1/2} \cdots$, apply \ref{Store}, split it back, and close $\dequeInv$.
	\item The rest is the same as the case without resizing, but uses $\gamma_{era-new}$, $arr_{new}$, and $L_{new}$.
\end{itemize}

\subsection{Steal}

In the full verification of \verb|steal|, we open the invariant five times: (1) reading the top, (2) reading the bottom, (3) loading the array, (4) reading the value at the top, and (5) CAS-ing the top. Compared to \autoref{chap:verif-simple}, there is an extra opening due to (3): loading $arr$ from $C \mapsto^{1/2} \cdots$ requires opening the invariant since we cannot duplicate it and keep it locally. Again, let $\gamma_{era-i}$, $arr_i$, $t_i$, $b_i$, and $L_i$ be the era, array pointer, top, bottom, and array, respectively, when $\dequeInv$ is opened for the $i$-th time.

Apart from that, the main difference in verification happens in (4). We open the invariant which gives $arr_4 \mapsto^{1/2} L_4$, but we want to read a value from the array we got in (3), so what we actually need is $arr_3 \mapsto^{1/2} \cdots$. We resolve this problem by a case analysis on whether the array was replaced in the meantime, i.e. $\gamma_{era-3} = \gamma_{era-4}$. If they are the same era, we use \ref{Dqst-Frag-Agree} to prove $arr_3 = arr_4$, and the rest is similar to \autoref{chap:verif-simple}. Otherwise, $arr_3$ is a past array, so we retrieve a points-to for it from $\dqstauth$ using \ref{Dqst-Archived-Get}. Then later, we use \ref{Dqst-Archived-Frag-Valid} to prove that the top has not increased if CAS succeeded.

The following is the proof detail for the case $\gamma_{era-3} \neq \gamma_{era-4}$:
\begin{itemize}
	\item We are currently at \verb|"v" := !(access (Fst "circle") "t" "sz")|.
	So far, we opened $\dequeInv$ for the 4th time,
	took a snapshot for each era, and
	know $t_1 \leq t_2 \leq t_3 \wedge \preservation{2}{3} \wedge \gamma_{era-3} \neq \gamma_{era-4}$.
	\\ (1) Use \ref{Dqst-Archived-Get} to get $\dqstarchf{}{\gamma_{era-3}, arr_3, L_A, t_A, b_A} \ast arr_3 \mapsto L'$.
	\\ (2) Use \ref{Dqst-Archived-Frag-Valid} to prove $t_3 \leq t_A \wedge \preservation{3}{A}$.
	\\ (3) Use \ref{Dqst-Archived-Frag-Get} to get $\dqstfragf{}{\gamma_{era-3}, arr_3, L_A, t_A, b_A}$.
	\\ (4) Use \ref{Dqst-Frag-Agree} to prove $|L_3| = |L'|$, and thus the value \verb|"sz"| we read for $L_3$ can still be used to read from $L'$.
	\\ (5) Read $v := L'[t_1]$ from $arr_3 \mapsto L'$.
	\\ (6) Return $\dqstarch$ back to $\dqstauth$. 
	\\ (7) Use \ref{Dqst-Frag-Valid} to prove $t_A \leq t_4 \wedge \preservation{A}{4}$.
	\\ (8) Close $\dequeInv$.
	\item \verb|CAS (top "deque") "t" ("t" + #1)|: open $\dequeInv$, use $\dqstauth$ to prove $t_4 \leq t_5$ by \ref{Dqst-Frag-Valid}, and attempt CAS with $top \mapsto t_5$.
	\item If $t_1 = t_5$, we succeed the CAS and we get $top \mapsto t_1+1$.
	Similarly to \autoref{chap:verif-simple}, prove $t_1 = t_2 = t_3 = t_A = t_4 = t_5$, $t_2 < b_2$, $t_3 < b_3$, $\cdots$, $t_5 < b_5$, and $L_5[t_5] = v$. Open $\au$, update the ghost states, commit, and close $\dequeInv$.
	\item If $t_1 \neq t_5$, we fail the CAS. Open $\au$, immediately commit, and close $\dequeInv$.
	\item End up with $\Phi$ and finish the proof.
\end{itemize}

\section{Construction of the Full Deque State}

To extend the definition of deque state in \autoref{chap:verif-simple} to support resizing, we need two key changes: (1) we need $\dqstarch$ to represent the last moment of an era. In particular, if we have both $\dqstfrag$ and $\dqstarch$ for the same era, we should be able to prove that $\dqstarch$ was made later than $\dqstfrag$. (2) $\dqstarch$ and the corresponding points-to to the array should be stored in $\dqstauth$ for each era.

\begin{figure}
\begin{mathpar}

\dqstauthf{\gamma}{\gamma_{era}, arr, L, t, b} :=
\begin{array}{l}
  \exists \gamma_{tb}, \gamma_{elt}, \gamma_{room}, M_{t}, M_{e}.
  \\
  \bigsep \left\{
    \begin{array}{l}
    \gamma = (\gamma_{tb}, \gamma_{elt}, \gamma_{room})
    \\
    \natauth{\gamma_{tb}}{tbs(t, b)} \ast \natauth{\gamma_{era}}{tbs(t, b)}
    \\
    \mapauth{\gamma_{elt}}{M_t} \ast
    (t = b \vee \mapfrag{\gamma_{elt}}{t}{L[t]})
    \\
	\forall k \in \dom(M_t). k \leq \begin{cases}t & \text{if }t < b \\ t-1 & \text{otherwise}\end{cases}
	\\
	\gamma_{era} \notin \dom(M_e)
	\\
	\mapauth{\gamma_{room}}{M_e[\gamma_{era} \leftarrow (arr, |L|)]} \ast
	\mapfrag{\gamma_{room}}{\gamma_{era}}{(arr, |L|)}
	\\
	\bigsep_{(\gamma' \mapsto (arr', n')) \in M_e}
		\exists L', t', b'.
		\dqstarchf{\gamma}{\gamma', arr', L', t', b'} \ast
		arr' \mapsto L'
    \end{array}
  \right.
\end{array}

\dqstfragf{\gamma}{\gamma_{era}, arr, L, t, b} :=
\begin{array}{l}
  \exists \gamma_{tb}, \gamma_{elt}, \gamma_{room}.
  \\
  \bigsep \left\{
    \begin{array}{l}
    \gamma = (\gamma_{tb}, \gamma_{elt}, \gamma_{room})
    \\
    \natfrag{\gamma_{tb}}{tbs(t, b)} \ast \natfrag{\gamma_{era}}{tbs(t, b)}
    \\
    t = b \vee \mapfrag{\gamma_{elt}}{t}{L[t]}
	\\
	\mapfrag{\gamma_{room}}{\gamma_{era}}{(arr, |L|)}
    \end{array}
  \right.
\end{array}

\dqstarchf{\gamma}{\gamma_{era}, arr, L, t, b} :=
\natauth{\gamma_{era}}{tbs(t, b)} \ast \dqstfragf{\gamma}{\gamma_{era}, arr, L, t, b}
\end{mathpar}

\caption{The definition of full deque state.}
\label{fig:deque_state_full_def}
\end{figure}

The full definition of deque state is given in \autoref{fig:deque_state_full_def}. To represent (1), the deque state has an additional copy of the monotonic natural number identified by $\gamma_{era}$; this is the purpose of representing each era as a ghost name. The numbers for $\gamma_{tb}$ and $\gamma_{era}$ stay identical in each era, but when we replace the array, we put the latter one into $\dqstarch$ and allocate a new monotonic natural number. To represent (2), the deque state also tracks an additional ghost map identified by $\gamma_{room}$. This map maps each era to the array pointer and the length of the array, representing the fact that they do not change within each array. Furthermore, for each mapping except for the current era, $\dqstauth$ stores an $\dqstarch$ and the points-to to the array.

The proof of each rule in \autoref{fig:deque_state_rule} is almost the same, and the additional rules in \autoref{fig:deque_state_more_rule} are proven as follows:
\begin{itemize}
	\item \ref{Dqst-Archived-Frag-Get}: trivial.
	\item \ref{Dqst-Archived-Frag-Valid}: use \ref{Mono-Nat-Lb-Valid} and \ref{Ghost-Map-Elem-Agree}, similarly to how we proved \ref{Dqst-Frag-Valid}.
	\item \ref{Dqst-Frag-Agree}: use \ref{Ghost-Map-Elem-Agree}.
	\item \ref{Dqst-Archived-Get}: use \ref{Ghost-Map-Lookup} to prove
	$M_e[\gamma_2 \leftarrow (arr, |L|)][\gamma_1] = (arr_1, |L_1|)$.
	Since $\gamma_1 \neq \gamma_2$, we get $(\gamma_1 \mapsto (arr_1, |L_1|)) \in M_e$.
	Use it to access $\dqstarchf{}{\gamma_1, arr_1, L, t, b} \ast arr_1 \mapsto L$
	from the separating conjunction $\bigsep_{(\gamma' \mapsto (arr', n')) \in M_e}$ of $\dqstauth$.
	\item \ref{Dqst-Archive}:
	\\ (1) Prove $(t = b \vee \mapfrag{\gamma_{elt}}{t}{L[t]})$ is preserved using $L[t..b) = L'[t..b)$.
	\\ (2) Take a snapshot $\dqstfragf{}{\gamma, arr, L, t, b}$ by \ref{Dqst-Frag-Get} and combine with $\natauth{\gamma}{tbs(t, b)}$ to get $\dqstarchf{}{\gamma, arr, L, t, b}$.
	\\ (3) Allocate a new $\natauth{\gamma'}{tbs(t, b)}$ for a new ghost name $\gamma'$ by a stronger variant of \ref{Mono-Nat-Alloc}: when making a new ghost name $\gamma'$, we can make sure that $\gamma'$ is different from all ghost names in some finite set. In our case, we ensure $\gamma' \notin \dom(M_e) \cup \{ \gamma \}$.
	\\ (4) Update the authoritative ghost map to
	$\mapauth{\gamma_{room}}{M_e [\gamma \leftarrow (arr, |L|)] [\gamma' \leftarrow (arr', |L'|)] }$ by \ref{Ghost-Map-Insert}
	and obtain $\mapfrag{\gamma_{room}}{\gamma'}{(arr', |L'|)}$.
	\\ (5) Prove the separating conjunction
	$\bigsep_{ (\gamma'' \mapsto (arr'', n'')) \in M_e [\gamma \leftarrow (arr, |L|)] }$
	of the new $\dqstauth$ by combining the prior $\bigsep$ with $\dqstarchf{}{\gamma, arr, L, t, b}$.
\end{itemize}

%% file: extended.tex
\label{chap:extended}
\section{Verification with Safe Memory Reclamation (SMR)}

In this section, we introduce a modular verification framework for memory reclamation schemes \cite{smr} and verify the Chase-Lev deque with memory reclamation.

\subsection{Background: Memory Reclamation}

To prevent memory leak in the programs manipulating memory, unused parts of the memory must be reclaimed. In the case of the Chase-Lev deque, after resizing the array, the older array must be deallocated. However, it should not be deallocated \textit{right away}: stealers might still be reading from the older array, resulting in a use-after-free bug. Deallocating is safe only after all stealers having access to the older array finish accessing it.

This can be resolved by keeping track of stealers for each array, so that the last stealer deallocates it. Of course, this method requires a large amount of additional code and verification. In the other direction, garbage collectors can be used to reclaim inaccessible memory automatically. However, it is unavailable in some low-level systems and languages, and incurs high performance overhead.

Various memory reclamation schemes \cite{hazard-pointer, ebr} have been proposed to strike a balance between simplicity and performance overhead. These schemes generally provide two APIs: (1) a function to \textit{protect} a part of the memory to prevent it from being deallocated; and (2) a function to \textit{retire} it so that it is deallocated when no threads are protecting it. This approach allows fine-grained memory control, and hides the detail of memory management at the same time.

\begin{figure}
  \noindent\begin{minipage}{\textwidth}
  \begin{lstlisting}[frame=tlrb,language=Coq]{Name}
Definition push : val := λ: "deque" "v", 
  ...
  (if: "t" + "sz" ≤ "b" + #1
    then let: "domain" := !("deque" +ₗ #qdom) in
      "deque" +ₗ #circle <- circle_grow "circle" "t" "b" "sz" ;;
      hazptr.(hazard_domain_retire) "domain" "circle" ("sz" + #1)
    else #()
  ) ;;
  ...

Definition deque_steal : val := λ: "deque",
  let: "domain" := !("deque" +ₗ #qdom) in
  let: "shield" := hazptr.(shield_new) "domain" in
  let: "t" := !("deque" +ₗ #qtop) in
  let: "b" := !("deque" +ₗ #qbot) in
  let: "circle" := hazptr.(shield_protect) "shield" ("deque" +ₗ #circle) in
  ...
  else let: "v" := !(circ_access ("circle" +ₗ #carr) "t" "sz") in
  hazptr.(shield_drop) "shield" ;;
  ...
  \end{lstlisting}
  \end{minipage}\hfill
  \caption{Changes in the implementation to support hazard pointers.}
  \label{fig:mod}
\end{figure}

\autoref{fig:mod} shows the modification necessary to support hazard pointers in Chase-Lev deque. In \verb|push|, after resizing and replacing the array, we retire the old array so that it can be deallocated when no stealers can access it. In \verb|steal|, we protect the circular array before accessing it, and drop the protection after the access. To port the code to the SMR verification framework, there is also a slight modification to how a deque is represented: the fields are now arranged in memory layout instead of a tuple.

\subsection{Specification and Verification in SMR}

To verify Chase-Lev deque under SMR schemes, the whole SMR scheme should be verified, and its specification should be incorporated into the verification of deque. The work by \citet{smr} fulfills this goal: it provides a modular specification for SMR in Iris. The specification is designed so that it can be seamlessly integrated into the data structures' verification without exposing the implementation details. The SMR schemes themselves have been verified in that work, so all we need is modify our Chase-Lev verification to use SMR specifications.

\begin{figure}
  \input{figure-hp-spec.tex}
  \caption{Selected proof rules of hazard pointers.}
  \label{fig:hp-spec}
\end{figure}

The specification of hazard pointers introduce three predicates:
(1) $\Managed (\ell, \gamma, P)$ representing the ownership of the pointer $\ell \mapsto \cdots$ managed by hazard pointers;
(2) $\Shield (s)$ representing a shield $s$ which can be used to protect a pointer; and
(3) $\Protected (s, \ell, \gamma, P)$ representing the protection of $\ell \mapsto \cdots$ by a shield $s$.
The predicate $P(\ell, v, \gamma)$ is a per-location invariant used for reasoning with its contents; it is usually a ghost state with the ghost name $\gamma$.

\autoref{fig:hp-spec} shows a few proof rules of hazard pointers. We can turn a points-to into a $\Managed$, which can be used to read a value and is discarded by calling \verb|hazard_domain_retire|. $\Managed$ also allows protecting the pointer with a shield $\Shield$, which turns it into a $\Protected$. This $\Protected$ can also be used to read a value, even after $\Managed$ is retired; the actual pointer is not deallocated until all protection to it is removed.

To combine the new SMR-related resources with Chase-Lev deque verification, we make the following changes from \autoref{chap:verif}:
\begin{itemize}
	\item Since the fields of the deque are arranged in memory, we change tuples to lists, e.g. for $(arr, |L|)$, we swap the two fields and concatenate them into $[|L|] + L$.
	\item We add $\gamma_{hp}$ for each era, a ghost name for the ghost variable to manage hazard pointers. This is stored in $M_e$ in addition to the array and its length.
	\item In $\dqstauth$, the points-to $arr \mapsto L$ for each past era no longer works since the array might have been deallocated. We replace it with $P(C, L, \gamma_{hp})$ where $P$ is a ghost variable: $P(\ell, v, \gamma) := \ghostVar{\gamma}{1/2}{v}$. Note that $P$ does not actually use $\ell$ in our proof.
	\item In $\dequeInv$ and $\ownDeque$, we extend the ghost variable for $\gamma_{era}$ to handle $\gamma_{hp}$.
	\item In $\dequeInv$, we replace $C \mapsto^{1/2} [|L|] + L$ with $\Managed(C, \gamma_{hp}, P) \ast P(C, L, \gamma_{hp})$.
	\item In $\ownDeque$, we remove $C \mapsto^{1/2} [|L|] + L$. The owner can still keep track of the array contents $L$ using the ghost variable for $\gamma_{sw}$.
	\item Now accessing the array requires $\Managed$ which is in $\dequeInv$, so we change the specification of \verb|grow| to a LAT. The atomic precondition and postcondition are $\Managed(C, \gamma_{hp}, P) \ast P(C, L, \gamma_{hp})$, and the private postcondition is $C' \mapsto [|L'|] + L'$.
\end{itemize}

For verification, we just need to apply the specification for SMR functions, and read the array using \ref{Managed-Access} and \ref{Protected-Access}. When the owner wants to read the array, we open $\dequeInv$ to get $\Managed(C, \gamma_{hp}, P) \ast P(C, L, \gamma_{hp})$. Then we apply \ref{Managed-Access} which gives $P(C, L', \gamma_{hp})$, and use \ref{Ghost-Var-Agree} on the two $P$ resources to prove $L = L'$.

For the stealer, we obtain $\Protected(s, C, \gamma_{hp}, P)$ when loading (and protecting) the array for the first time. Later, we read the array from $\Protected$ and proceed again by case analysis on whether the era has advanced before CAS-ing. If not advanced, we follow the same procedure as above. Otherwise, we retrieve the $\gamma_{hp}$ of the past era from $\dqstauth$ and use it to agree on the array.

\section{Foundation for Verification in Relaxed Memory Model}

In this section, we briefly discuss the verification of Chase-Lev deque in relaxed memory model. Due to the complexity of this memory model, we leave the full specification and verification to future work. Instead, we show how to verify the safety of Chase-Lev deque, which is expected to be required for full verification.

\subsection{Background: Relaxed Memory Model and iRC11}

In relaxed memory model, instructions can be reordered as long as the overall behavior is preserved. For example, adjacent instructions \verb|X <- 1| and \verb|Y <- 1| can be swapped since they do not affect each other. However, this kind of reordering allows more behaviors in the context of concurrency. Consider the following example, where all variables are initially 0 and two codes separated by a line are executed concurrently:

\begin{tabular}{l||l}
     \verb|X = 1| & \verb|if Y == 1| \\
     \verb|Y = 1| & \verb|    assert(X == 1)|
\end{tabular}

Unlike the SC memory model, it is possible to fail the assertion in the relaxed memory model: the code on the left is reordered and \verb|Y <- 1| is executed, and then we enter the \verb|Y == 1| branch even though \verb|X == 1| does not hold.\footnote{In order to ensure correctness for codes like this, a certain ordering can be enforced using release-acquire ordering, memory fences, and so on, but they are beyond the scope of this paper.}

Instead of directly accounting for reordering, we adopt the equivalent view-based semantics \cite{promising, gps, hai-thesis}. Here, the instructions are executed in order, but they may read past values. Specifically, each thread maintains their \textit{views}, the read and write events it has observed so far. When reading a value from a location, the thread can read a write event for that location that it has not observed yet. Similarly, when writing to a location, the thread can insert a write event after the last one it has observed.

\textit{iRC11} \cite{rbrlx, hai-thesis} is a separation logic for relaxed memory, formalized in Iris. It incorporates the idea of views into the logic by maintaining multiple events in each location. It introduces a points-to assertion of the form $\ell \mapsto h$ where $h$ is a \textit{history}, a list of events at the location $\ell$, which can be used to read one of the events of $h$. An event is composed of a timestamp at which the event was made, a value written and a view tied into it. The timestamp is used to determine whether the thread can read the event. Upon reading an event, the thread gets the value as the result of reading and combines the view with the thread's own view.

To simplify the logic further, iRC11 supports the following access modes: (1) \textit{single-writer} mode, in which only one thread can write to the location and all other threads can only read; (2) \textit{CAS-only} mode, in which all threads can write to the location but only via CAS-ing; and (3) \textit{read-only} mode, in which no threads can write to the location. As the interface of these modes, the following resources are defined: \begin{itemize}
	\item $\ell \mapsto_\SW h$, $\ell \mapsto_\CAS h$, the points-to assertions with the history $h$, denoting that $\ell$ is being used in the single-writer and CAS-only mode respectively;
	\item $\ell \mapsto_\RO n$, the read-only points-to assertion with the value (not history) $n$;
	\item $\ell \snseen h$, the \textit{history-seen observation}, asserting that the thread has observed all write events in the history $h$;
	\item $\ell \swseen h$, the \textit{single-writer ownership}, asserting that the thread has an exclusive right to write to $\ell$, with the full history being $h$ which has been completely observed;
	\item $\ell \roseen n$, the \textit{read-only observation}, the right to read $n$ from $\ell \mapsto_\RO n$.
\end{itemize}

Unlike the points-to in \autoref{chap:iris}, the points-to in iRC11's specialized access modes cannot be split. Instead of checking for the fractions to read or write, it additionally requires an assertion about observation. To read from or write to a location $\ell$, the thread must at least own $\ell \mapsto_\theta h$, where $\theta$ is either \SW \ or \CAS, and $h$ is a history. In addition, for writing in the single-writer mode, the thread must own $\ell \swseen h$. For the other operations (reading in any mode, or CAS-ing in the CAS-only mode), the thread must own $\ell \snseen h'$ for some history $h'$.

\subsection{Implementation and Safety Verification in iRC11}

The implementation of Chase-Lev deque in relaxed memory model is almost the same as the SC model, except that we should appropriately insert memory fences and determine the ordering mode of each operation. However, the choice of fences and ordering does not affect safety, only functional correctness, so we skip the details here and just follow the implementation by Kang \cite{cl-weakmem-lin}.


We eventually would like to verify a strong specification with Compass \cite{compass}, but for now we only verify that each operation runs safely:
\begin{mathpar}
	\hoare{0 < n}{new\_deque(n)}{ p. \exists \gamma.
		\knowInv{}{\dequeInvg(p)} \ast \ownDequeg(p) \ast \dequeLocal^\gamma(p)
	}
	\\
	\knowInv{}{\dequeInv(p)} \vdash \dequeLocal \wand
	\hoare{\ownDeque(p)}{push(p, v)}{\ownDeque(p)}
	\\
	\knowInv{}{\dequeInv(p)} \vdash \dequeLocal \wand
	\hoare{\ownDeque(p)}{pop(p)}{\_. \ownDeque(p)}
	\\
	\knowInv{}{\dequeInv(p)} \vdash \dequeLocal \wand
	\hoare{\TRUE}{steal(p)}{\_. \TRUE}
\end{mathpar}

Here, $\dequeLocal$ is a persistent resource each thread maintains locally. It asserts that the thread has observed ($\snseen$) some history for the top, bottom, and array, so that it can read from the corresponding points-to.

To verify the specifications above, we put points-to, history observation, and ownership into $\dequeInv$, $\ownDeque$, $\dequeLocal$ appropriately. This time, each era is represented by the timestamp $\TS$ at which the array was written to the array pointer. We maintain two ghost maps: persistent one for array pointer and the length of the array, and non-persistent one for the list $H$ of histories for the array, i.e. the list for which each $H[i]$ is the history for the $i$-th slot of the array. These ghost maps are used for synchronization between $\dequeInv$ and $\ownDeque$.

Now we define the resources necessary for the specification. First, $\dequeInv$ stores $\mapsto_\SW$ for the array and bottom, $\mapsto_\CAS$ for the top, and $\mapsto_\SW$ and $\snseen$ for each of the archived array:

$$
\dequeInvg(p) :=
\begin{array}{l}
	\exists \gamma_{era}, \gamma_{arr}, C, top, bot, h_t, h_b, h_C.
	\\
	\bigsep \left\{
\begin{array}{l}
    \gamma = (\gamma_{era}, \gamma_{arr})
    \wedge
    p = (C, top, bot)
    \\
	C \mapsto_\SW h_C
	\wedge top \mapsto_\CAS h_t
	\wedge bot \mapsto_\SW h_b
	\\
	\bigsep_{\TS \mapsto (p_\TS, V_\TS)}
	\circleInv(\TS, p_\TS, V_\TS, \gamma_{era}, \gamma_{arr})
\end{array}
	\right.
\end{array}
$$

$$
\circleInv(\TS, p, V, \gamma_{era}, \gamma_{arr}) :=
\begin{array}{l}
	\exists arr, H.
	\\
	\bigsep \left\{
\begin{array}{l}
	p = (arr, |H|) \wedge |H| > 0
	\\
	\mapfrag{\gamma_{era}}{\TS}{(p, |H|)} \ast \mapfraghalf{\gamma_{arr}}{\TS}{H}
	\\
	arr \mapsto_\SW H
	\\
	\bigsep_{arr[i] = h} \exists h'. h' \subseteq h \ast arr \snseen h'
\end{array}
	\right.
\end{array}
$$

Next, $\ownDeque$ stores $\swseen$ for the array, bottom, and each slot of the current array. It also asserts that all values ever written to the bottom are larger than 0, and the current array is the last one written to the history of the array pointer:

$$
\ownDequeg(p) :=
\begin{array}{l}
	\exists \TS_{last}, H_{last}, M_e, M_a, \gamma_{era}, \gamma_{arr}, C, top, bot, arr, h_b, h_C.
	\\
	\bigsep \left\{
\begin{array}{l}
	\gamma = (\gamma_{era}, \gamma_{arr}) \wedge p = (C, top, bot)
	\\
	\dom(M_e) = \dom(M_a) = \dom(h_C)
	\\
	bot \swseen h_b \ast C \swseen h_C
	\\
	\bigsep_{\TS \mapsto (b, \_) \in h_b} b \geq 1
	\\
	\TS_{last} = \max(\dom(h_C)) \wedge h_C[\TS_{last}] = (arr, \_)
	\\
	\bigsep_{arr[i] = h} arr[i] \swseen h
	\\
	\mapauth{\gamma_{era}}{M_e} \ast \mapauth{\gamma_{arr}}{M_a}
	\\
	\mapfrag{\gamma_{era}}{\TS}{(arr, |H_{last}|)} \ast \mapfraghalf{\gamma_{arr}}{\TS}{H_{last}}
\end{array}
	\right.
\end{array}
$$

Finally, $\dequeLocal$ stores $\snseen$ for the array, top, and bottom:

$$
\dequeLocal^\gamma(p) :=
\begin{array}{l}
	\exists \gamma_{era}, \gamma_{arr}, h_t, h_b, h_C.
	\\
	\bigsep \left\{
\begin{array}{l}
	\gamma = (\gamma_{era}, \gamma_{arr}) \wedge p = (C, top, bot)
	\\
	top \snseen h_t \ast bot \snseen h_b \ast C \snseen h_C
\end{array}
	\right.
\end{array}
$$

The verification of each operation is not very interesting: it is just a matter of opening the invariant, using \ref{Ghost-Var-Agree} to unify values if necessary, using $\mapsto$ and $\snseen$ or $\swseen$ to read and write, and closing the invariant. However, the formalized proof is still very long. For comparison: the Coq formalization of the full verification in \autoref{chap:verif} involves complicated reasoning about synchronization, the definition of deque state resources (e.g. $\dqstauth$), all proof rules for the deque state along with the proof of each rule. Despite that, it is only slightly longer than the safety proof in iRC11 which skips all synchronization and doesn't even define the deque state resources.

%% file: figure-hp-spec.tex
\begin{mathpar}
  \axiomH{Managed-New}
  {\ell \mapsto v \ast P(\ell, v, \gamma) \vsW \Managed(\ell, \gamma, P)}
  \and
  \axiomH{HP-Retire}
  {\hoare
    {\Managed(\ell, \gamma, P)}
	  {retire(\ell)}
    {\TRUE}}
  \\
  \axiomH{Managed-Access}
  {\hoare
    {\Managed(\ell, \gamma, P)}
	{! \ell}
    {\Ret v. P(\ell, v, \gamma) \ast \Managed(\ell, \gamma, P)}}
  \\
  \axiomH{Protect}
  {\ahoare
    {src \mapsto \ell * \Managed(\ell, \gamma, P) * \Shield(s)}
	{protect(s, src)}
    {\Ret \ell. src \mapsto \ell * \Managed(\ell, \gamma, P) * \Protected(s, \ell, \gamma, P)}}
  \\
  \axiomH{Protected-Access}
  {\hoare
    {\Protected(s, \ell, \gamma, P)}
	{! \ell}
    {\Ret v. P(\ell, v, \gamma) \ast \Protected(s, \ell, \gamma, P)}}
\end{mathpar}

%% file: conclusion.tex
\label{chap:conclusion}
\section{Summary}

We have formally verified Chase-Lev deque using the Iris separation logic.
This is the first known verification ofthe  Chase-Lev deque that is foundational, uses a realistic and unbounded implementation, and verifies a strong specification.
We also extended the verification to incorporate safe memory reclamation techniques, and established a basis for verifying the deque in the relaxed memory model.

\section{Related Work}

While various papers have introduced new work-stealing deque designs, most of them lack formal proofs or rely solely on pen-and-paper proofs to establish their correctness.
Implementations without formal verification pose a risk of containing errors, even after an extensive testing, as evidenced by bug reports in commercial softwares.
Chase-Lev deque is no exception to this issue.
\citet{cl-weakmem} proposed ARM and C11 implementations of Chase-Lev deque in relaxed memory model, and proved the ARM implementation correct.
However, the C11 implementation lacked a formal proof and was later discovered to have a bug \cite{cdschecker}.
Also, although a proof can increase confidence in correctness to some degree, pen-and-paper proofs are still prone to human mistake and require thorough review.
We can further instill confidence by checking the proof computationally, including our work.

However, prior works on mechanized verification of work-stealing deques have limitations like weaker specification, restrictive or unrealistic implementations, and larger trusted computing base, compared to our work.
Here, we list such prior works and outline their limitations.

\citet{abp-mc} verified the linearizability of the ABP work-stealing deque \cite{abp} using model checking techniques.
However, they made the simplifying assumption of an infinitely large array, which is not realistic in practice.
Moreover, the ABP deque itself has an inherent limitation due to its bounded capacity; in fact, Chase-Lev deque was specifically designed to address this limitation \cite{cl}.
Also, their verification method was not fully mechanized.
Instead of directly encoding linearizability into model checking, they checked a few basic properties and proved on pen-and-paper that they together imply linearizability.

In a later work, \citet{genmc} developed GenMC, a model checker for C programs under configurable memory models.
As a part of the benchmark, they verified Chase-Lev deque in relaxed memory model.
However, their implementation of the deque has a bounded capacity.
Although it uses a circular array, it does not dynamically resize, so the owner thread fails to push its tasks when the array is full.
Also, the verification targeted a weaker specification instead of linearizability \cite{genmc-bench-cl}.

\citet{bwos} proposed a novel block-based work-stealing deque, and verified its correctness using GenMC.
Similarly to the above work, they did not verify strong specifications, but only weaker guarantees like each element being popped or stolen only once.

\citet{cl-refine} took a different approach to verify Chase-Lev deque by using layered refinement.
This approach consists of multiple layers of implementation.
Starting from the target implementation, the code is gradually transformed into a simpler versions that refine the previous ones.
Eventually each operation becomes physically atomic, at which point logical atomicity of the target implementation is proven since it is refined by its physically atomic form.
However, their implementation assumed an infinitely large array which is again unrealistic.
This also led to skipping the resizing procedure, which would simplify the synchronization reasoning compared to the real implementaion.

Finally, while the aforementioned works have benefits of automation, they do not achieve foundational verification.
The correctness of the whole verification process relies on trusting the verification tools themselves.
For instance, although GenMC \cite{genmc} has been formalized and proven to be sound and complete, the proof is pen-and-paper and the correctness of its C++ implementation is not guaranteed either.
Similarly, the work by \citet{cl-refine} uses the Civl verifier \cite{civl}, which is a complex proof system consisting of multiple steps:
it is built upon another verifier that incorporates an SMT solver, and Civl introduces its own processor as well.

Outside of work-stealing deques, there is another example of a foundationally verified scheduling queue.
\citet{dartino} verified the Dartino queue, a scheduling data structure used in Google's Dartino virtual machine, in the Iris separation logic.
As Dartino queue is implemented as a lock-based linked list, it has a simpler synchronization reasoning compared to Chase-Lev deque.
The invariant of the Dartino queue mainly consists of standard properties related to node ownership.

\section{Future Work}

Verifying the full specification of Chase-Lev deque in the relaxed memory model poses two challenges.
First, Chase-Lev deque utilizes SC memory fences, which are currently not supported by iRC11.
Second, unlike the SC memory model, there is no widely accepted strong specification for data structures in the relaxed memory model \cite{compass}.
Simple linearizability and logical atomicity may not be suitable for some relaxed-memory data structures such as Herlihy-Wing queue \cite{linearizability} and exchanger \cite{exchanger},
because instructions in the relaxed memory model may only be synchronized with a subset of other instructions.
The difficulty of designing a strong specification further complicates the verification.

To address the first point, we plan to extend iRC11 to support SC memory fences, in a way that the synchronization guarantees achieved by the fences are adequately modeled.
Regarding the second point, \citet{compass} developed Compass, a framework for strong specifications of relaxed memory data structures built on top of Iris.
It offers comprehensive support for functional correctness properties, including synchronization and FIFO guarantee for queues.
Moreover, it enables modular client reasoning for these data structures.
However, the specification and proof are long and complicated, in part due to the inherent complexity of the relaxed memory.
To address this gap, we are working on designing a simpler interface to faciliate easier verification.
In addition, we are exploring the extension of Diaframe \cite{diaframe}, an Iris proof automation tool, to automate parts of the verification process for the relaxed memory model.

Furthermore, we plan to extend the verification of safe memory reclamation schemes \cite{smr} to the relaxed memory model.
Combined with the Compass framework, we aim to achieve verified strong specifications of various data structures, including Chase-Lev deque.

In the other direction, it would be interesting to verify more recent designs of work-stealing deques.
Various techniques have been proposed and demonstrated to improve synchronization overheads, such as
private deques \cite{acar, lace},
architecture-aware optimizations \cite{fence-free},
relaxation of stealing guarantees \cite{idempotent, fully-fence-free},
or block-based task grouping \cite{bwos}.
Verifying these data structures would require more sophisticated reasoning and invariants, especially considering the novel techniques they employ.

Furthermore, the verification of work-stealing deques can be extended to verify schedulers that make use of these deques.
While there have been extensive research efforts in the area of verified schedulers and operating systems \cite{certikos, sel4},
to the best of our knowledge, support for work-stealing scheduling schemes is not yet present.
It would be intriguing to explore how the whole work-stealing strategy can be verified using the specifications for work-stealing deques.


%% file: ack.tex
\acknowledgment[4]

저의 석사 과정을 처음부터 끝까지 이끌어주신 강지훈 교수님께 깊은 감사의 말씀을 드립니다. 늘 부족한 제가 연구를 잘할 수 있도록 지도해 주시고, 미래에 대해 걱정이 될 때마다 귀중한 조언을 해주셨습니다. 교수님의 도움 없이는 저의 연구가 세상에 나오지 못했을 것입니다.

저의 곁에 있어주신 KAIST 동시성 및 병렬성 연구실의 모든 구성원 분들께 감사드립니다. 특히 석사 과정 전부터 지금까지 동시성 검증 프로젝트를 함께 진행하며 제 연구의 방향성에 대해 큰 도움을 주신 정재황 님과 이장건 님, 그리고 느슨한 메모리 검증 프로젝트를 함께해 주신 박선호 님과 김재우 님께 감사드립니다.

마지막으로, 저의 평생을 지켜보며 응원해 주시는 저의 가족에게 감사드립니다. 저의 결정을 늘 존중하며 아낌없는 지원을 해주신 부모님, 제 삶의 4년 선배로서 늘 귀감이 되어준 형에게 감사합니다. 덕분에 제가 이 자리에 서게 될 수 있었습니다.